\documentclass[12pt]{iopart}

\expandafter\let\csname equation*\endcsname\relax
\expandafter\let\csname endequation*\endcsname\relax

\usepackage{dcolumn}
\usepackage{bm}
\usepackage{graphicx}
\usepackage{pifont}
\usepackage{revsymb}
\usepackage{array}
\usepackage{booktabs}
\usepackage{mathtools}
\usepackage{graphics}
\usepackage{dsfont} 
\usepackage{amsmath,amssymb}
\usepackage{hyperref}
\usepackage{tabularx}
\usepackage{epsf,epsfig}
\usepackage[normalem]{ulem}
\usepackage[usenames]{color}
\usepackage{multirow}
\usepackage{makecell}
\usepackage{diagbox}
\usepackage[abs]{overpic}
\usepackage[dvipsnames]{xcolor}
\allowdisplaybreaks
\hypersetup{
    colorlinks=true,
    linkcolor=blue,
    filecolor=magenta,      
    urlcolor=blue,
    citecolor=blue
}
\newcommand{\iWD}{{\mbox{\tiny WD,in}}}
\newcommand{\oWD}{{\mbox{\tiny WD,out}}}
\newcommand{\PSR}{{\mbox{\tiny PSR}}}
\newcommand{\NS}{{\mbox{\tiny NS}}}
\newcommand{\BH}{{\mbox{\tiny BH}}}
\newcommand{\EdGB}{{\mbox{\tiny EdGB}}}
\newcommand{\GB}{{\mbox{\tiny GB}}}

\newcommand{\DEF}{{\mbox{\tiny DEF}}}

\definecolor{red(ncs)}{rgb}{0.77, 0.01, 0.2}
\definecolor{blue(ncs)}{rgb}{0.0, 0.01, 0.77}

\usepackage[perpage]{footmisc}

\begin{document}

\title[Probing Scalar-Tensor Theories with Gravitational Waves from Mixed Binaries]{Future Prospects for Probing Scalar-Tensor Theories with Gravitational Waves from Mixed Binaries}

\author{Zack Carson}
\address{Department of Physics, University of Virginia, Charlottesville, Virginia 22904, USA}

\author{Brian C. Seymour}
\address{Department of Physics, University of Virginia, Charlottesville, Virginia 22904, USA}

\author{Kent Yagi}
\address{Department of Physics, University of Virginia, Charlottesville, Virginia 22904, USA}

\date{\today}


\begin{abstract} 
The extreme-gravity collisions of binaries with one black hole and one neutron star provide for excellent tests of general relativity. We here study how well one can constrain theories beyond general relativity with additional scalar fields that allow for spontaneous scalarization of neutron stars, and those motivated from string theory. We find that existing bounds can be improved with current gravitational-wave detectors if the black hole mass is sufficiently small. Bounds will further improve by many orders of magnitude with future detections, especially by combining multiple events. 
\end{abstract}

\maketitle


\section{Introduction}\label{sec:intro}
Einstein's famous theory of general relativity (GR) describes the relationship between matter, and the curvature of spacetime through a single tensor, the metric $g_{\mu\nu}$.
Over the last 100 years, several alternative theories of gravity have been proposed -- yet countless observations and tests of GR in a variety of environments have proven to be consistent with Einstein's theory: solar system~\cite{Will_SolarSystemTest}, pulsar timing~\cite{Stairs_BinaryPulsarTest,Wex_BinaryPulsarTest} and cosmological~\cite{Ferreira_CosmologyTest,Clifton_CosmologyTest,Joyce_CosmologyTest,Koyama_CosmologyTest,Salvatelli_CosmologyTest} observations.
With such a rigorous upholding to all of our tests, why should we bother to further prove or disprove the solid theory of GR?
The answer lies within the many unanswered questions we still have today, obscuring our understanding of the universe.
Some examples of these include the unification of GR and quantum mechanics, the inflationary period in the early universe, dark matter and its influence on the galactic rotation curves, and dark energy and its impact on the expansion of our universe.
Alternative theories of gravity, if found to be true, could explain the missing links to our unanswered questions.

More recently, a new observational opportunity into the extreme-gravity (strong, non-linear, and highly-dynamical) regime~\cite{Abbott_IMRcon2,Yunes_ModifiedPhysics} has been unveiled with the recent gravitational wave (GW) detections of coalescing black holes (BHs)~\cite{GW150914,LIGOScientific:2018mvr} and neutron stars (NSs)~\cite{TheLIGOScientific:2017qsa}.
To date, eleven GW observations have been made~\cite{LIGOScientific:2018mvr}, and have further found no statistically significant deviations from GR~\cite{Abbott_IMRcon,Monitor:2017mdv,Abbott:2018lct}.

In this analysis, we consider the present and future implications on constraining non-GR theories with an additional massless scalar field, known as scalar tensor theories (STTs), using both GWs and pulsar timing observations.
The latter has been studied for binary pulsar systems~\cite{Wex_BinaryPulsarTest,Anderson:2019eay}, pulsar-white dwarf (WD) systems~\cite{Freire:2012mg,Wex_BinaryPulsarTest,Shao:2017gwu,Anderson:2019eay}, and triple pulsar-WD-WD systems~\cite{Berti_ModifiedReviewLarge,Archibald:2018oxs}.
Here, we consider the present and future constraints obtained from the GW detections of BH-NS coalescences.
See also Refs.~\cite{Bonilla:2019mbm,DAgostino:2019hvh} for constraints on STTs from GWs.
In STTs, compact objects acquire scalar charges that source the scalar field. A scalar force acts between two scalarized objects, giving rise to a fifth force which depends on the internal structure of the massive objects and violates the strong equivalence principle (SEP). Binaries consisting of scalarized astrophysical objects further emit scalar dipole radiation (on top of gravitational quadrupolar radiation in GR), causing the binaries to evolve faster.  

Such radiation becomes larger when the difference between the scalar charges of the binary constituents become larger, and thus a mixed binary consisting of one black hole and one neutron star system is ideal for probing such theories~\cite{Berti:spaceFreq,Takahiro,Sagunski:2017nzb,Huang:2018pbu}. 
Specifically, the increased mass difference $m_1^2-m_2^2$ and small total mass $M$ of a BH/NS system will minimize the allowable constraints on $\sqrt{\alpha}\sim M$ (for non-GR coupling parameter $\alpha$), while the offset acquired from the increased SNR of the alternative scenario of a large-mass binary BH system\footnote{In many cases, low-mass binary BH systems and BH/NS systems with slowly rotating BHs may be indistinguishable. However, in the Einstein-dilaton Gauss-Bonnet theory of gravity, the dipole-radiation slightly decreases  when comparing a BH/NS system with the equivalent binary BH system with the NS replaced by a slowly rotating BH but the effect is insignificant, and thus the constraints are not significantly affected.} only improves constraints by a small factor of $\sim \text{SNR}^{-1/4}$, while the SNR only itself increases by $\sim M^{5/6}$. Moreover, a smaller binary system has a lower relative velocity for a fixed frequency, which leads to an enhanced dipole radiation.
It is then extremely advantageous to decrease the total mass of the system rather than maximize it.
Such sources are particularly interesting and extremely timely to consider as two of the candidates in the O3 run by the LIGO/Virgo Collaborations, S190426c and S190814bv, are likely to be the merger of a black hole and a neutron star, if they are of astrophysical origin~\cite{gracedb,gracedb2}\footnote{We note that the LIGO/Virgo Collaborations categorize BH/NS candidate events as $m_1>5\text{ M}_\odot$ and $m_2<3\text{ M}_\odot$~\cite{classification}, and therefore O3 candidates S190426c and S190814bv potentially could be binary BH mergers. In addition, S190426c currently has a 58\% possibility of being terrestrial noise.}.
Because BHs have vanishing scalar charges in the quasi-Brans-Dicke theory of gravity considered in this analysis, the presence of a NS is required to place constraints on such a theory.
Therefore, it is of vital importance that, not only the events be of astrophysical origin, but they must with high confidence also be a BH/NS system rather than a binary BH system.

\begin{figure}
\begin{center}
\includegraphics[width=.7\columnwidth]{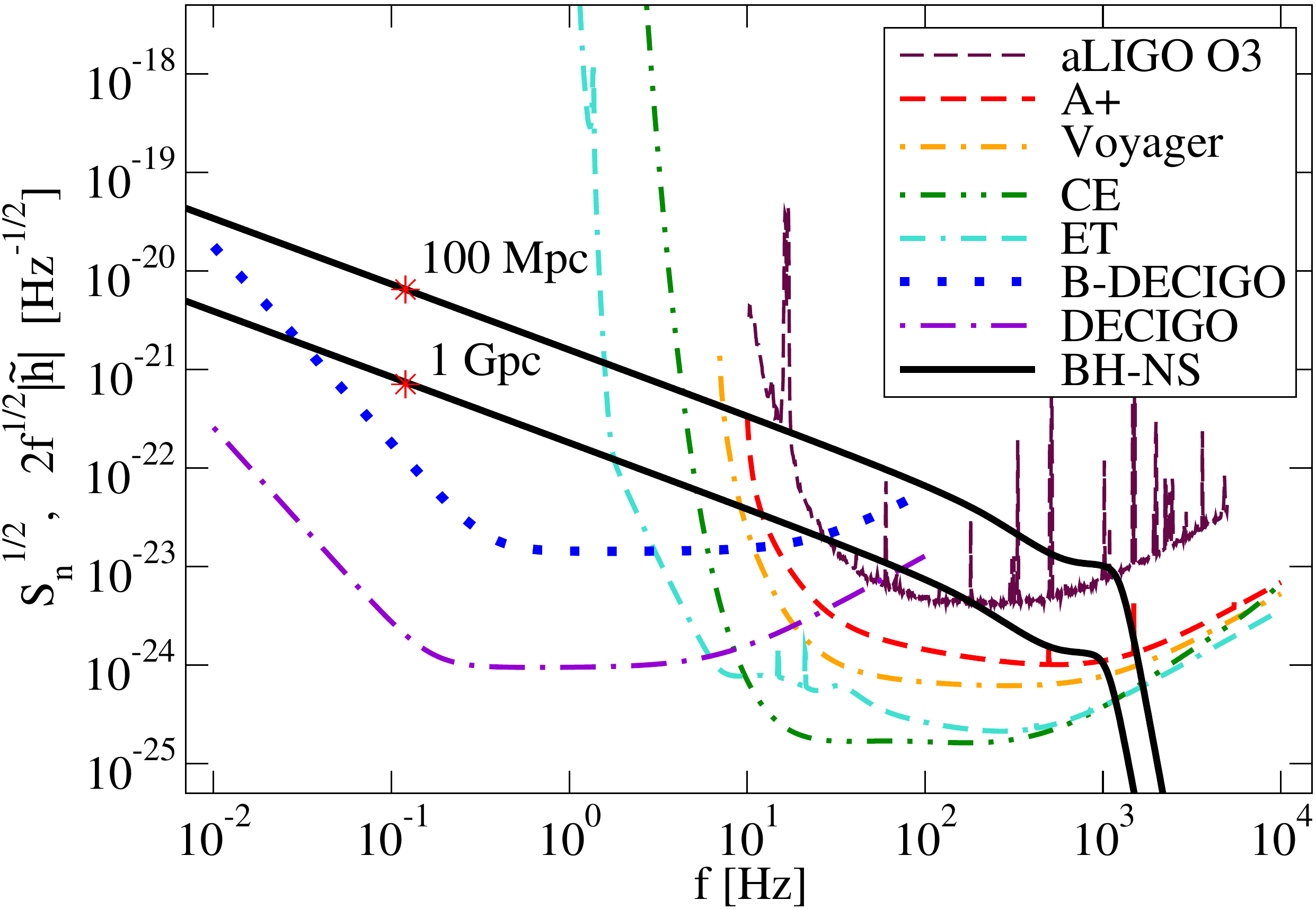}
\caption{
Detector sensitivities $\sqrt{S_n(f)}$ for the various interferometers considered in this analysis.
Additionally shown are the characteristic amplitudes $2\sqrt{f}|\tilde{h}(f)|$ for a $10\text{ M}_\odot-1.4\text{ M}_\odot$ BH-NS systems $100$ Mpc and $1$ Gpc away, where the latter is not detectable by aLIGO O3.
The red stars indicate one year prior to merger, corresponding to the lower integration limit for space-based detectors.
In the above curves, the ratio between the sensitivity and the characteristic amplitude roughly corresponds to the SNR of the event.
}\label{fig:noiseCurves}
\end{center}
\end{figure}

A particularly interesting class of theories within STTs are quasi-Brans-Dicke theory and Einstein-dilaton Gauss-Bonnet (EdGB) gravity. The former was introduced first by Damour and Esposito-Far$\acute{e}$se (DEF)~\cite{Damour:1992we,Damour:1996ke} which induces a non-linear growth of the scalar charges onto neutron stars called \emph{spontaneous scalarization }~\cite{Damour:1992we,Damour:1993hw,Damour:1996ke}, while black holes remain hairless as in GR. The latter is motivated from string theory and the dilaton scalar field is coupled with a quadratic curvature term (Gauss-Bonnet invariant) in the gravitational action~\cite{Kanti_EdGB,Maeda:2009uy}. In this theory, black holes have non-vanishing scalar charges~\cite{Campbell:1991kz,Yunes:2011we,Yagi_EdGBmap,Sotiriou:2014pfa} while neutron stars do not if the scalar field coupling is linear~\cite{Yagi_EdGBmap,Yagi:2015oca}.
We here consider the single BH-NS detections with future GW detectors, as well as the multi-band detections between both space- and ground-based detectors~\cite{Barausse:2016eii,Carson_multiBandPRL,Nair:2015bga,Nair:2018bxj}, and finally the combination of multiple observations~\cite{Takahiro,Abadie:2010cf} made on future detectors with expanded horizons.

The outline of the paper is as follows.
We begin in Sec.~\ref{sec:techniques} with a discussion on 
the gravitational waveform model, GW interferometers, Fisher analysis techniques used in our analysis, together with information on combining uncertainties from multiple events.
In Sec.~\ref{sec:theory}, we review the theoretical background on STTs and present our main results.
Section~\ref{sec:approximations} provides an analysis and description of the validity of the various approximations made in our investigation.
Finally, we conclude and offer avenues of future direction in Sec.~\ref{sec:conclusion}.
Throughout this paper, we have adopted the geometric units of $G=1=c$.


\section{Non-GR gravitational waveforms and data analysis}\label{sec:techniques}
Let us begin by considering how to capture modifications to GR in the gravitational waveform from compact binary mergers.
Typically, one strives to be agnostic towards the multitudes of alternative theories of gravity available, by modifying the phase of the GR waveform by $\Psi_{\text{GR}}+\Delta\Psi$, where $\Delta\Psi\equiv \beta u^{2n-5}$.
Here, $u\equiv (\pi \mathcal{M} f)^{1/3}$ is the effective relative velocity of binary constituents with GW frequency $f$ and chirp mass $\mathcal{M}\equiv(m_1^3m_2^3/M)^{1/5}$ with individual masses $m_A$ and total mass $M=m_1+m_2$.
The post-Newtonian (PN) parameter\footnote{An $n$PN-order term is proportional to $(u/c)^{2n}$ relative to the leading-order term in the waveform.} $n$ categorizes the order at which the modifications affect the GW phase, and $\beta$ prescribes the magnitude of the modification.
This \emph{parameterized post-Einsteinian} (ppE) formalism~\cite{Yunes:2009ke} allows for one to effectively constrain any modified theory of gravity by knowing the ppE expression $\beta$ (which can be mapped to the coupling parameters of SEP-violating theories) and the PN order $n$ at which the leading correction enters the waveform, many of which are tabulated in e.g.~\cite{Tahura_GdotMap}.

Let us start with the detector sensitivity curves utilized in this analysis: aLIGO (representative of observing run O3)~\cite{aLIGO,O3} (O3), LIGO A\texttt{+}~\cite{AppData,ApVoyagerCE}, LIGO Voyager~\cite{ApVoyagerCE,VRTCEETData}, Cosmic Explorer~\cite{ApVoyagerCE,Evans:2016mbw,VRTCEETData} (CE, standard configuration described in Fig.~1 of~\cite{Evans:2016mbw}), Einstein Telescope configuration D~\cite{ApVoyagerCE,VRTCEETData,Evans:2016mbw} (ET), B-DECIGO~\cite{B-DECIGO}, and DECIGO~\cite{DECIGO}.
Figure~\ref{fig:noiseCurves} shows the resulting sensitivity curves of each interferometer, as well as the characteristic amplitudes $2\sqrt{f}|\tilde{h}(f)|$ for $10\text{ M}_\odot-1.4\text{ M}_\odot$ BH-NS systems located $100$ Mpc and $1$ Gpc away.
Here, it can be seen that only the former system can be detected by the O3 detector.

We implement the sky-averaged IMRPhenomD waveform model (such that the inclination angle and sky-location parameters are averaged over) found in Refs.~\cite{PhenomDII,PhenomDI}.
Because accurate BH/NS tidal waveform models do not yet exist to sufficient accuracy for the purposes of this investigation\footnote{See Ref.~\cite{Lackey:2013axa}, where a phenomenological BH/NS waveform model was constructed where the phase exceeded the NR results by 30\%. See also Ref.~\cite{Kumar:2016zlj}, which updates the model from Ref.~\cite{Lackey:2013axa} with a more accurate baseline binary BH model, and Ref.~\cite{Pannarale:2015jka} for a BH/NS amplitude model, Ref.~\cite{Hinderer:2016eia} for an effective-one-body model applicable to BH/NS systems, or Ref.~\cite{Barkett:2019tus} for BH/NS models computed with tidal splicing. Finally, refer to Ref.~\cite{Chakravarti:2018uyi} for an analysis on the waveform systematic uncertainties present in such models.}, we modify the waveform with the simple $5$PN+$6$PN finite size tidal corrections for the NS~\cite{Wade:tidalCorrections}.
However, it is unlikely that such tidal effects (first entering at 5PN) are vital in the constraint of STT parameters (first entering at -1PN) due to minuscule correlations between the two.
Additionally, we do not consider the effects of spin precession.
See Sec.~\ref{sec:rotating} for a discussion on the inclusion of this effect.
We also note that in this waveform model, there is no contribution from the finite-size effect to the spin-induced quadrupole moment, first entering at $2$PN order. 
However, this is likely to make a negligible effect in our results due to the small spin priors assumed on the NS, as described below.

To estimate constraints on $\beta$ from future GW observations, we utilize a Fisher analysis as described in Ref.~\cite{Cutler:Fisher}.
Assuming sufficiently loud signals and Gaussian-distributed noise, Fisher-techniques allow one to predict the root-mean-square error on waveform template parameters $\theta^a$ to be $\Delta\theta^a\approx\sqrt{(\Gamma^{-1})^{aa}}$.
Here, $\bm{\Gamma}$ is the Fisher information matrix detailed in Eq. (2.7) of Ref.~\cite{Cutler:Fisher}, and depends on the detector sensitivity and waveform model.
The waveform model used has template parameters of luminosity distance $D_L$, masses $m_A$ and spins $\chi_A$ of the BH and NS (See~\ref{app:EosSpinCompare} for a demonstration of the importance of including spin-effects in the waveform), the time and phase at coalescence $t_c$ and $\phi_c$, NS tidal deformability parameter $\Lambda$\footnote{The tidal deformability of a non-rotating BH is zero~\cite{damour-nagar,Binnington:2009bb,Kol:2011vg,Chakrabarti:2013lua,Gurlebeck:2015xpa}. See also Ref.~\cite{Gralla:2017djj} in which it was shown that the standard computation of the tidal deformability relies on the comparison to the BH one and the effective BH tidal deformability may have a small non-zero effect on the gravitational waveform.}, and the ppE parameter $\beta$.
We assume Gaussian priors on the spins to be $|\chi_\BH|<1$ and $|\chi_\NS|<0.05$~\cite{Harry:2018hke}, and on the tidal parameter of $0<\Lambda<5000$~\cite{LIGO:posterior}\footnote{Such priors are constructed by translating the (non-Gaussian) values quoted above into the 68\% confidence intervals of Gaussian probability distributions.}.
The Fisher analysis technique described here is only used an approximation to the more comprehensive (and computationally expensive) Bayesian analysis. However Ref.~\cite{Yunes:2016jcc} showed that bounds on ppE parameters sufficiently agree between the two methods for GW150914 with SNR$\sim24$.
Thus, we expect the events considered in this analysis to give sufficiently valid order-of-magnitude estimations.

The upper and lower frequency integration limits used for each detector in the Fisher analyses are tabulated in Table~\ref{tab:freqLimits}.
For ground-based detectors, the upper cutoff frequency of $f_{\text{ISCO}}=(6^{3/2}\pi M_z)^{-1}$ with redshifted mass $M_z\equiv(1+z)M$ is used for each ground-based detector, with lower cutoff frequencies of 10 Hz, 10 Hz, 5 Hz, and 5 Hz for A\texttt{+}, Voyager, CE and ET respectively\footnote{Reference~\cite{Evans:2016mbw} mentions a starting frequency of 10 Hz for CE, yet the noise curve extends down to 5 Hz and is thus used here for completeness (such an extension is not made for A\texttt{+} and Voyager, which also extend to $5$ Hz.). Additionally, very little SNR is accumulated between 5-10 Hz and thus the signal between $5-10$ Hz is expected to have a negligible impact on our results. This was indeed tested, and it was found that the difference was negligible to the accuracy of our results.}.
The redshift is computed in a cosmology such that $D_L=\frac{1+z}{H_0}\int^z_0\frac{dz'}{(1-\Omega_\Lambda)(1+z')^3+\Omega_\Lambda}$ with Hubble constant $H_0=70$ km s$^{-1}$ Mpc$^{-1}$ and vacuum energy density $\Omega_\Lambda=0.67$.
Following Ref.~\cite{Berti:spaceFreq}, the upper cutoff frequencies for space-based detectors is taken to be 100 Hz, with lower cutoff frequencies corresponding to one year prior to merger.
We note that we use the redshifted chirp mass in the computation of such frequencies.

\begin{table}
\centering
\begin{tabular}{c|c|c}
Detector & $f_{\text{low}}$ [Hz] & $f_{\text{high}}$ [Hz]\\
\hline
A\texttt{+} & 10 & 386 \\
Voyager & 10 & 386 \\
CE &  5 & 386 \\
ET &  5 & 386 \\
B-DECIGO & 0.12 & 100 \\
DECIGO & 0.12 & 100 \\
\end{tabular}
\caption{Upper and lower frequency limits used in the Fisher matrix integrations of Ref.~\cite{Cutler:Fisher}.
For ground-based detectors the upper cutoff frequency is determined to be $f_{\text{ISCO}}$ for a BH-NS binary with masses $10M_\odot$ and $1.4M_\odot$, while space-based detectors use lower cutoff frequencies determined to be one year prior to merger~\cite{Berti:spaceFreq}.
}\label{tab:freqLimits}
\end{table}

Additionally shown in Table~\ref{tab:eventRates} are the one-year BH-NS detection rates used in our analysis to combine uncertainties, for each detector considered.
Such rates are computed following Ref.~\cite{Takahiro} as
\begin{equation}
N=\Delta T\int\limits_0^{z_h}4\pi \lbrack a_0 r(z) \rbrack^2 \mathcal{R} R(z) \frac{d\tau}{dz}dz,
\end{equation}
where $\Delta T=1$ yr is the observing time, $z_h$ is the horizon distance\footnote{The Horizon distance is defined as the luminosity distance such that an threshold SNR of $8$ is observed.} redshift of each detector, $\mathcal{R} \in \lbrack 0.6, 610 \rbrack \text{Gpc}^{-3}\text{yr}^{-1}$ (with a ``realistic value" of $30 Gpc^{-3}yr^{-1}$) is the local BH-NS coalescence rate~\cite{LIGOScientific:2018mvr,Abadie:2010cf}, and $a_0r(z)$, $d\tau/dz$, and $R(z)$ can all be found in Ref.~\cite{Takahiro}.
Finally, the combined uncertainty on a parameter $\theta^a$ can be computed as
\begin{equation}\label{eq:stacking}
\sigma_{\theta^a}^{-2}=\Delta T\int\limits_0^{z_h}4\pi \lbrack a_0 r(z) \rbrack^2 \mathcal{R} R(z) \frac{d\tau}{dz} \sigma_{\theta^a}(z)^{-2} dz,
\end{equation}
where $\sigma_{\theta^a}(z)$ is the root-mean-square error on $\theta^a$ as a function of redshift, computed via Fisher analyses for increasing luminosity distances $D_L$.
See Sec.~\ref{sec:varymass} for an alternative analysis in which the BH and NS masses are varied for a more comprehensive combined uncertainty calculation.

\begin{table}
\centering
\begin{tabular}{|c|c|c|c|}
\hline
\multirow{2}{*}{Detector} & \multicolumn{3}{c|}{Detection Rate} \\
\cline{2-3}\cline{3-4}
& Pessimistic & Realistic & Optimistic\\
\hline
\hline
A\texttt{+}&5&270&5,500\\
Voyager&72&3,600&74,000\\
CE&720&36,000&730,000\\
ET&510&25,000&520,000\\
B-DECIGO&43&2,200&44,000\\
DECIGO&730&37,000&730,000\\
\hline
\end{tabular}
\caption{Pessimistic, mean, and optimistic 1 year detection rates for $10\text{ M}_\odot-1.4\text{ M}_\odot$ BH-NS binaries assuming a local BH-NS coalescence rate of $\mathcal{R} \in \lbrack 0.6, 610 \rbrack \text{Gpc}^{-3}\text{yr}^{-1}$, with a ``realistic value" of $30\text{ Gpc}^{-3}\text{yr}^{-1}$~\cite{Abadie:2010cf,LIGOScientific:2018mvr}.
}\label{tab:eventRates}
\end{table}

\section{Scalar-tensor theories of gravity}\label{sec:theory}
In this section, we discuss the two primary STTs of gravity presented in the following analysis. 
These include the quasi-Brans-Dicke theory DEF, as well as the EdGB theories of gravity.

\subsection{Quasi-Brans-Dicke theories}\label{sec:bransdicke}
Let us first focus on a class of scalar-tensor theories, quasi-Brans-Dicke theory.
In this theory, matter fields couple to the scalar field $\varphi$ through the effective metric $A^2(\varphi)g_{\mu\nu}$~\cite{Shao:2017gwu,Damour:1993hw,Damour:1996ke,Anderson:2019hio}.
One can then define the gradient and curvature of the conformal potential $\ln A(\varphi)$ to be $\alpha(\varphi) \equiv d\ln{A(\varphi)}/d\varphi$, and $\beta(\varphi) \equiv d\alpha/d\varphi$.
In particular, we focus on the Damour and Esposito-Far$\acute{e}$se (DEF)~\cite{Damour:1996ke,Damour:1992we} theory\footnote{See~\ref{app:theoryCompare} for a comparison with the similar Mendes-Ortiz (MO)~\cite{Mendes:2016fby} model.}, where the coupling function can be written in one of its simplest forms as $A(\varphi)=\exp{(\beta_0 \varphi^2/2)}$.
Such a theory can be completely characterized by the two weak-field parameters $\alpha_0=\alpha(\varphi_0)=\beta_0\varphi_0$ and $\beta_0=\beta(\varphi_0)$, where $\varphi_0= \alpha_0/\beta_0$ is the asymptotic value of the scalar field $\varphi$ at infinity~\footnote{$\beta_0 < 0$ leads to cosmological runaway evolution of the scalar field that violates the current solar system bounds~\cite{Damour:1992kf,Sampson2014,Anderson:2016aoi}, unless one introduces a mass to the scalar field either directly or effectively by e.g. coupling the scalar field to an inflaton~\cite{Anson:2019ebp}.}.

Similarly in the strong-field case, NSs with mass $m_A$ couple to the scalar field with an effective coupling $\alpha_A=\partial \ln m_A/\partial\phi_0$, known as the (dimensionless) \emph{scalar charge} \footnote{Scalar charges depend on the NSs underlying equation-of-state (EoS). In this analysis we assume the APR4 EoS, consistent with the binary NS observation GW170817~\cite{LIGO:posterior,TheLIGOScientific:2017qsa}. See~\ref{app:EosSpinCompare} for a comparison between results found with different EoSs.} (the scalar charge for BHs is 0~\cite{Damour:1992we}).
Such scalar charges induce scalar dipole radiation in a compact binary, which enters at $-1$PN order relative to the leading GR quadrupole radiation and makes the binary evolve faster.
Following Ref.~\cite{Damour:1998jk,Shao:2017gwu,Littenberg:2018xxx}, one can derive the corresponding ppE correction to the waveform to be
\begin{equation}\label{eq:ppeBeta}
\beta_\DEF=-\frac{5\eta^{2/5}(\Delta\alpha)^2}{7168}, \hspace{5mm} n=-1,
\end{equation}
where $\eta\equiv m_1m_2/M^2$ is the symmetric mass ratio, and $\Delta\alpha\equiv(\alpha_1-\alpha_2)$ is the difference in scalar charges between orbiting compact objects.
Additionally, see Sec.~\ref{sec:highorder} for a discussion and comparison on the inclusion of higher-order PN corrections to the waveform phase, as well as to the amplitude.
See also Ref.~\cite{Zhao:2019suc} for constraints from GW170817, and predictions for future binary NS detections.

Now let us discuss how one can constrain STTs with pulsar timing measurements. The first way to do this is through orbital decay rate $\dot P_{b}$ measurement. The dominant correction to orbital decay rate in STTs is from the dipole radiation of the scalar field $\dot P_b^D$. Thus, we constrain $\dot P_b^D/\dot P_{\text{GR}}$ by the fractional measurement accuracy of the orbital decay rate $\delta_{\dot P_b}$. In STTs, the expression for $\dot P_b^D/\dot P_{\text{GR}}$ is
\begin{equation}\label{eq:pdotdipole}
    \frac{\dot P_b^D}{\dot P_{\text{GR}}} = \frac{5}{96} \frac{G}{1+\alpha_0^2}\frac{(\Omega_b m_2)^{-2/3}}{(1+q)^{2/3}}\frac{1+e^2/2}{1+\frac{73}{24}e^2+\frac{37}{96}e^4}(\Delta\alpha)^2 <\delta_{\dot P_b} \, ,
\end{equation}
where $\Omega_b$ is the orbital frequency, $m_2$ is the pulsar's companion mass, $q$ is the mass ratio $m_1/m_2$, and $e$ is the orbital eccentricity. Using the upper bound on dipole radiation in Eq.~\eqref{eq:pdotdipole}, we can place constraints on the theory parameters. The second way of testing STTs with pulsar timing is through constraints on the SEP. 
Recently, PSR J0337+1715 has placed the most stringent bounds on SEP violation~\cite{Archibald:2018oxs}. 
This SEP violation measurement $\Delta$ constrains the theory parameter of STTs with the inequality $|\alpha_\oWD(\alpha_\PSR-\alpha_\iWD)|<\Delta$ where $\alpha_\PSR$, $\alpha_\oWD$, and $\alpha_\iWD$ are the scalar charges of the pulsar, outer WD, and inner WD respectively.
We update the constraints on STTs from Ref.~\cite{Archibald:2018oxs} by using a softer EOS, because stiff EoSs are inconsistent with the recent GW observations~\cite{TheLIGOScientific:2017qsa,Abbott:LTposterior,LIGO:posterior}.

\begin{figure}
\begin{center}
\includegraphics[width=.7\columnwidth]{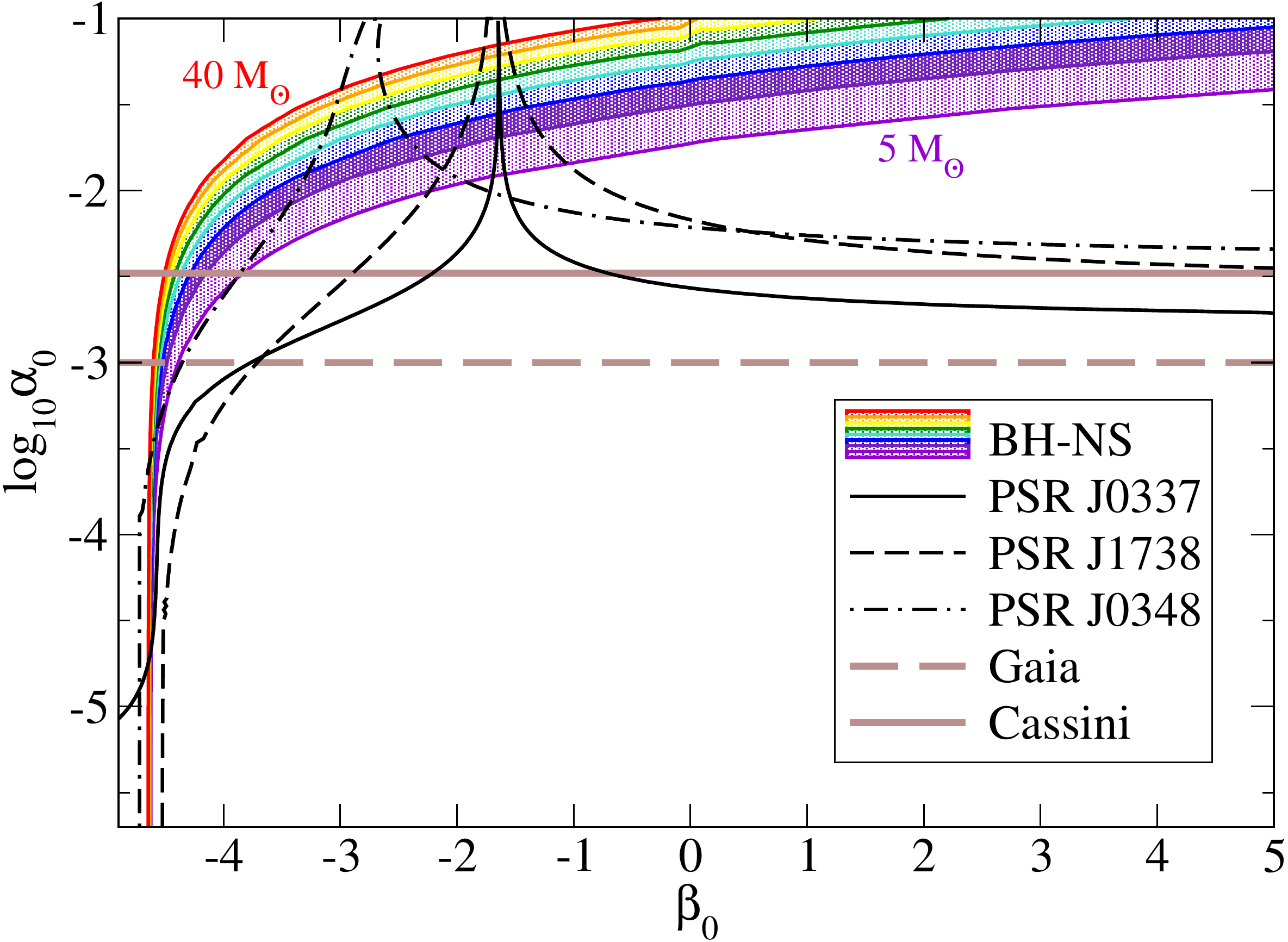}
\caption{Estimated $68$\% confidence interval bounds on the DEF quasi-Brans-Dicke modified theory of gravity with an assumed EoS of APR4, detected on the LIGO O3 detector with a SNR of 10 that is close to a detection threshold SNR and thus the bounds serve as conservative.
Such bounds are presented for a BH-NS system with $m_\NS=1.4\text{ M}_\odot$ and $m_\BH$ varying between $5\text{ M}_\odot$ and $40\text{ M}_\odot$ (with iterations of $5\text{ M}_\odot$).
The solid, dashed, and dash-dotted black curves correspond to constraints placed by the pulsar triple system PSR J0337+1715~\cite{Ransom:2014xla,Archibald:2018oxs}, and the pulsar-WD systems PSR J1738+0333~\cite{Kilic:2014yxa} and PSR J0348+0432~\cite{Antoniadis:2013pzd}, respectively.
The solid and dashed brown horizontal lines correspond to constraints by the existing Cassini spacecraft~\cite{Bertotti:2003rm} and those predicted by Gaia~\cite{mignard_2009}.
Such bounds are computed via $\alpha_0^2=\frac{|1-\gamma|}{2-|1-\gamma|}$ for parameterized post-Newtonian parameter $\gamma$ (see Eq.~(18) of Ref.~\cite{Anderson:2019eay}).
Take note that the Cassini constraints converted here to $\alpha_0$ were obtained with a few assumptions that make them applicable as an order-of-magnitude estimation.
}\label{fig:aligoDEF}
\end{center}
\end{figure} 

We first discuss the present considerations of DEF constraints using GW and pulsar timing observations.
Figure~\ref{fig:aligoDEF} presents the estimated constraints in the DEF theory parameter $\alpha_0-\beta_0$ plane for the various observations considered in this analysis.
Observe that the combination of Cassini and pulsar timing measurements from PSR J0337 and PSR J1738 places the strongest constraints on DEF gravity. Moreover, even if the O3 candidates S190426c and S190814bv were BH-NS merger events~\cite{gracedb}, they struggle to place competitive bounds on DEF theory. Thus, this motivates why we must consider future bounds on DEF from GW measurements.

We conclude with an expedition into the future of GW astronomy.
We consider the BH-NS system described previously, with fixed BH and NS masses of $10\text{ M}_\odot$ and $1.4\text{ M}_\odot$, respectively.
We assume detections on the future GW interferometers A\texttt{+}, Voyager, CE, ET, B-DECIGO, and DECIGO, and following the spirit of Ref.~\cite{Takahiro} we combine the bounds on $\Delta\alpha$ from $N$ BH-NS detections falling within the horizon of each detector over one observing year, as described in Sec.~\ref{sec:techniques}.
Further, we consider the multi-band observations~\cite{Carson_multiBandPRD,Carson_multiBandPRL,Gnocchi:2019jzp} of such binaries between both ground-based detector ET and space-based detectors DECIGO/B-DECIGO.
Unlike the multi-band case with space-based detector LISA, in which the possibly large SNR threshold of $\sim9$~\cite{Moore:2019pke} would prevent one from obtaining event rates larger than $\mathcal{O}(1)$, multi-band detections between ET and DECIGO/B-DECIGO will instead be limited by the ET event rate because of the large SNRs obtained by the space-based telescopes (this reasoning primarily applies to DECIGO, with event rates approximately equivalent or greater than those on ET and CE, rather than B-DECIGO with significantly smaller rates).
Such event rates are still significantly large at $\sim (500-500,000)$ (see Table~\ref{tab:eventRates}).

\begin{figure}
\begin{center}
\includegraphics[width=\textwidth]{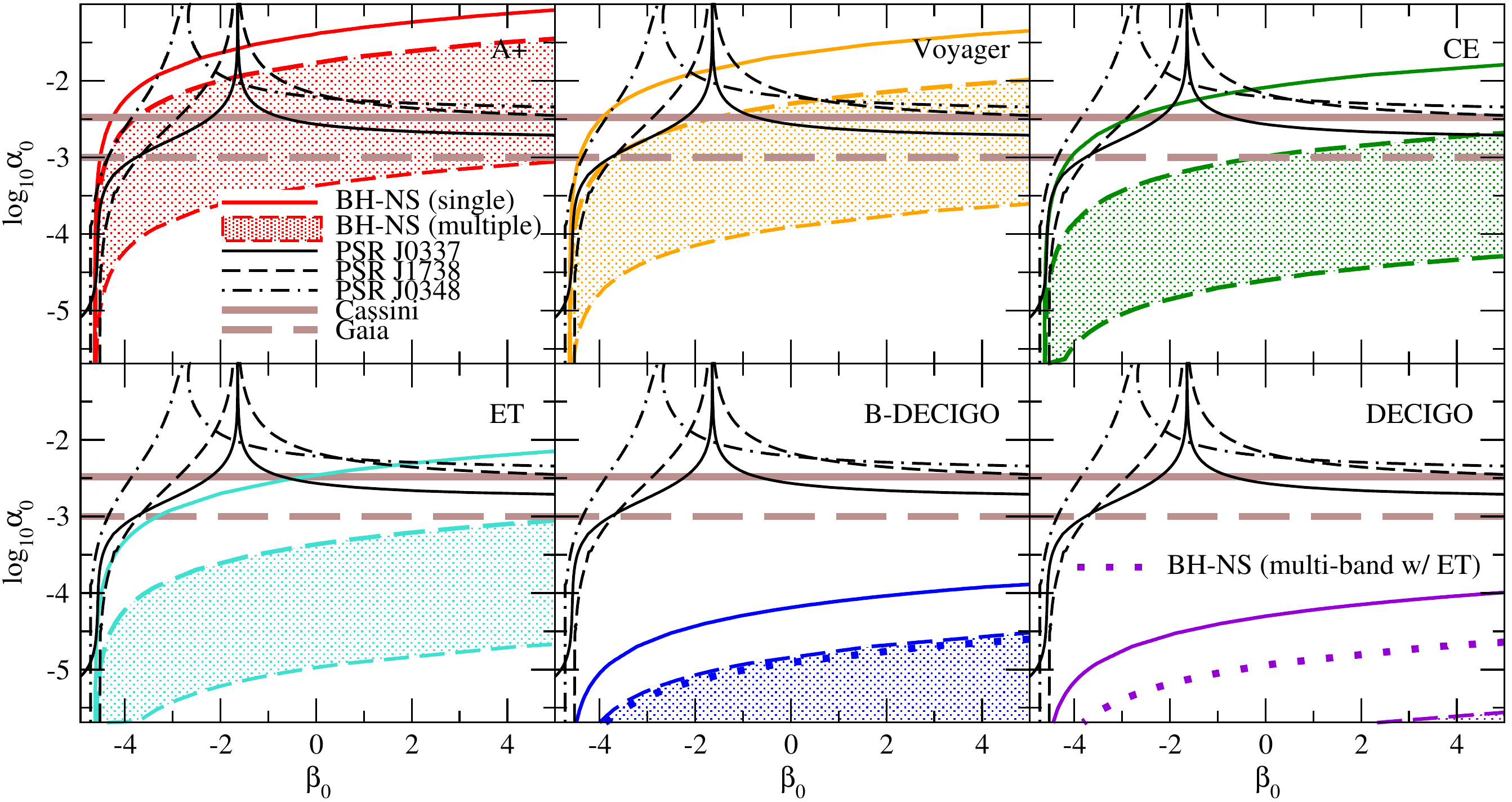}
\caption{Predicted $68$\% confidence interval constraints on the DEF quasi-Brans-Dicke modified theory of gravity with an assumed EoS of APR4, for $10\text{ M}_\odot-1.4\text{ M}_\odot$ BH-NS merger events at 1 Gpc detected by A\texttt{+}, Voyager, CE, ET, B-DECIGO, and DECIGO, with SNRs of 8.5, 21, 143, 71, 24, and 600 respectively.
In each panel, the solid colored lines represent bounds for single events, while the shaded region between dashed colored lines represent the combined constraints from multiple events, from the pessimistic to optimistic coalescence rates.
Additionally, the dotted lines represent the bounds from the multi-band observations between ET and B-DECIGO/DECIGO.
The brown horizontal lines and solid/dashed/dot-dashed black curves are the same as Fig.~\ref{fig:aligoDEF}.
}\label{fig:stackedDEF}
\end{center}
\end{figure}

Figure~\ref{fig:stackedDEF} presents the bounds in the DEF theory placed for the above-mentioned procedures.
Observe how all of the current constraints can be improved upon with the optimistic number of detections on the A\texttt{+} detector, while CE and ET begin to approach the same point with only the pessimistic number of detections.
Further, all predicted bounds placed with DECIGO/B-DECIGO (single-event, multiple-event, and multi-band) improve the current constraints by several orders of magnitude.
Of course, existing bounds from solar system experiments and binary pulsar observations will also improve in future. For example, bounds on $\alpha_0$ from Gaia will improve those from Cassini by a factor of a few~\cite{mignard_2009}, while the bounds from the pulsar triple system PSR J0337 will improve by a factor of $\sim 10$ with SKA~\cite{Berti_ModifiedReviewLarge}. Future GW bounds with 3rd generation detectors (ET/CE) and space-based detectors (B-DECIGO/DECIGO) are likely to be even stronger than them.
We also note that the bounds for Brans-Dicke theory with $\beta_0=0$ for ET and DECIGO are consistent with those in~\cite{Zhang:2017sym,Takahiro}.


\subsection{Einstein-dilaton Gauss-Bonnet gravity}\label{sec:EdGB} 
We now show how the GW observations of BH-NS binaries can be applied to constrain another alternative scalar-tensor theory: EdGB gravity.
In this string-inspired theory, the dilaton scalar field $\varphi$ non-minimally couples to a quadratic curvature term. In particular, we consider a linear coupling, where the Einstein-Hilbert action is corrected with a term $\alpha_\EdGB \varphi R^2_\GB$~\cite{Kanti_EdGB,Maeda:2009uy,Sotiriou:2013qea}\footnote{See Refs.~\cite{Bakopoulos:2018nui,Antoniou:2017hxj,Antoniou:2017acq} for general couplings.}.
Here, $\alpha_\EdGB$ is the coupling parameter of the theory while $R_\GB$ is the Gauss-Bonnet invariant.
In this SEP-violating theory of gravity, BHs can accumulate scalar charges~\cite{Campbell:1991kz,Yunes:2011we,Takahiro,Sotiriou:2014pfa} while ordinary stars like NSs do not~\cite{Yagi_EdGBmap,Yagi:2015oca}. Similar to quasi-Brans-Dicke theory, such charges will induce a scalar dipole radiation in a binary involving at least one BH, which accelerates the rate of inspiral between gravitating bodies.
Such an effect modifies the gravitational waveform phase at $-1$PN order, and is proportional to the coupling parameter $\alpha_\EdGB$ of the theory, as well as the masses $m_A$ and sensitivities $s_A$ of the compact bodies~\cite{Yagi_EdGBmap} (see Ref~\cite{Tahura_GdotMap} for the appropriate ppE expression).
The spin-dependent sensitivities are non-zero for BHs only, and are taken to be 1 (spinless) for this analysis.
The current constraints on EdGB gravity have been found to be $\sqrt{\alpha_\EdGB} \lesssim 2$ km~\cite{Yagi_EdGB,Nair_dCSMap,Yamada:2019zrb}\footnote{We note that the small-coupling approximation $\zeta_\EdGB\equiv16\pi\alpha_\EdGB^2/M^4 \ll 1$ for binaries with total mass $M$ must be satisfied in order for valid constraints on $\sqrt{\alpha_\EdGB}$ to be placed.}.
We urge caution that the constraint $\sqrt{\alpha_\EdGB} \lesssim 2$ found in Refs.~\cite{Nair_dCSMap,Yamada:2019zrb} take into account certain approximations which warrant such results to be a rough estimate.
For example, the authors of Ref.~\cite{Nair_dCSMap} used posterior samples provided by the LIGO/Virgo Collaborations though data more finely-sampled around the GR value seem to be necessary to derive a more reliable posterior distribution on $\sqrt{\alpha_\EdGB}$.
In addition, the authors utilize $-1$PN constraints obtained by the LIGO-Virgo analysis of the binary black hole signals, where it is warned that they cannot be interpreted as dipole radiation constraints, due to higher-order non-negligible terms from dipole radiation.
Also, Ref.~\cite{Yamada:2019zrb} utilized a simplified grid-search method rather than a full stochastic sampling procedure.

\begin{figure}
\begin{center}
\includegraphics[width=.7\columnwidth]{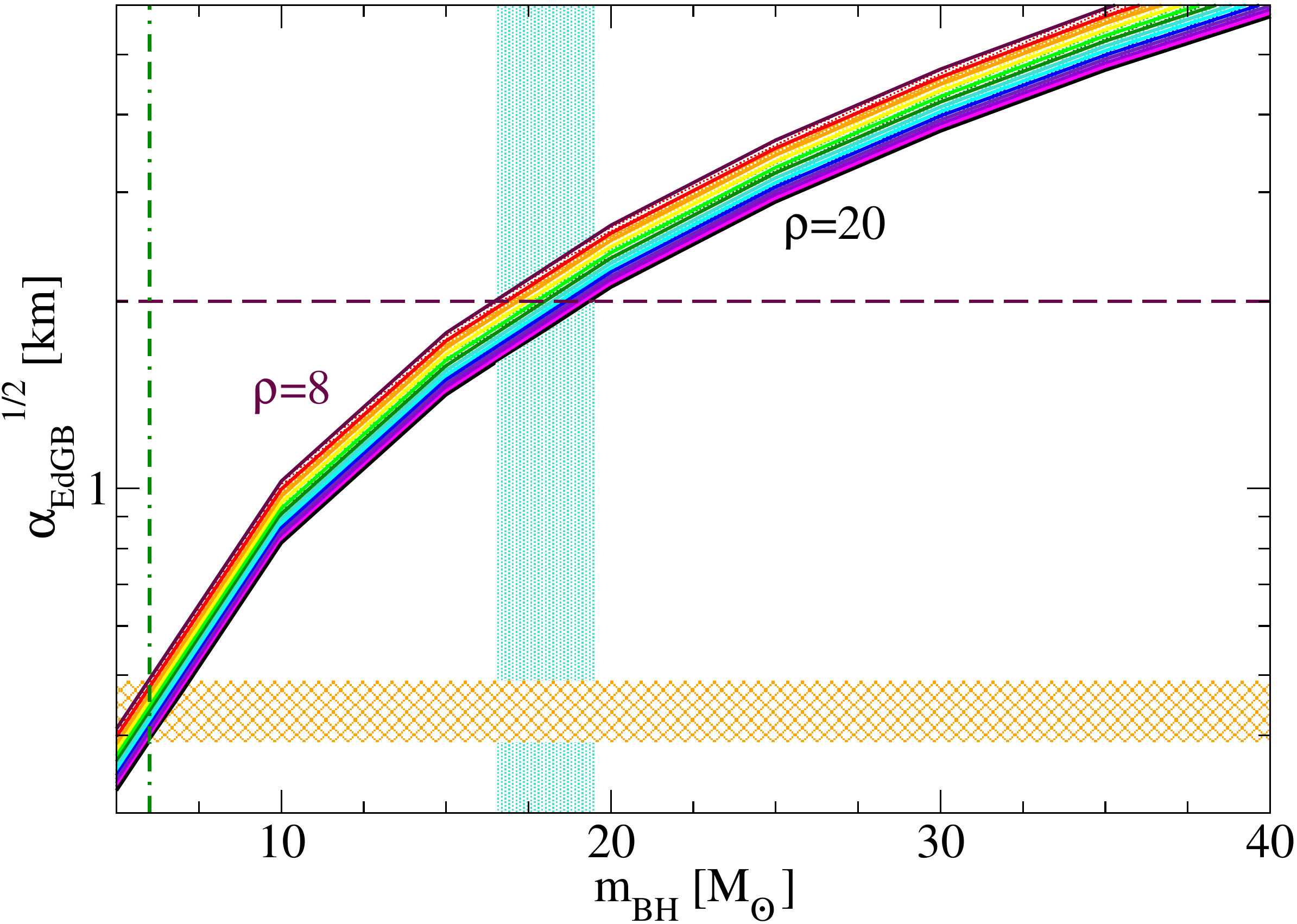}
\caption{Projected $68$\% confidence interval bounds on the EdGB coupling parameter $\sqrt{\alpha_\EdGB}$ as a function of the black hole mass $m_\BH$ merging with a $1.4\text{ M}_\odot$ NS.
Such constraints are presented for event SNRs ranging from $\rho=8$ to 20 with iterations of 1.
Observe that the strongest constraint in the literature~\cite{Yagi_EdGB,Nair_dCSMap,Yamada:2019zrb} can be improved upon for events with $m_\BH<16.5\text{ M}_\odot$, with the intersection displayed by the vertical shaded turquoise region.
Observe also that the BH/NS candidate S190426c (with a 58\% probability of terrestrial origin rather than astrophysical) with a likely BH mass of $\sim6\text{ M}_\odot$~\cite{Lattimer:2019qdc} can place a constraint of $\sqrt{\alpha_\EdGB}<\lbrack0.4-0.5\rbrack$ km, indicated by the shaded orange region, which is stronger than the current bound by a factor of 4--5.
}\label{fig:aligoEDGB}
\end{center}
\end{figure}

We begin by discussing the current observational constraints on $\sqrt{\alpha_\EdGB}$, had a BH-NS coalescence been observed by the current iteration of LIGO interferometers.
Figure~\ref{fig:aligoEDGB} projects the prospective constraints on $\sqrt{\alpha_\EdGB}$ for BH-NS binaries with $m_\NS=1.4\text{ M}_\odot$ as a function of $m_\BH$ for detection SNRs ranging between 8 and 20 on the aLIGO O3 detector.
Observe how for BHs with mass less than $16.5\text{ M}_\odot$ ($19.5\text{ M}_\odot$), the current constraint in the literature of $2$ km can be improved upon for events with SNR $=8$ (20).
Thus, if S190426c or S190814bv are NS-BH merger events with sufficiently low-mass BHs, such events would place a bound in EdGB gravity that is stronger than the existing bounds.
Reference~\cite{Lattimer:2019qdc} estimated the properties of S190426c from the probability of the system being in specific categories, such as BH/NS. 
In particular, the BH mass is estimated as $\sim 6M_\odot$. 
If S190426c was indeed a BH/NS system (58\% probability of terrestrial origin) and if this mass estimate was correct, we can place strong constraints on $\sqrt{\alpha_\EdGB}$ of $0.4$ ($0.5$) km for a 20 (8) SNR event -- a factor of 4--5 improvement from the current observational constraint.

\begin{figure}
\begin{center}
\includegraphics[width=.7\columnwidth]{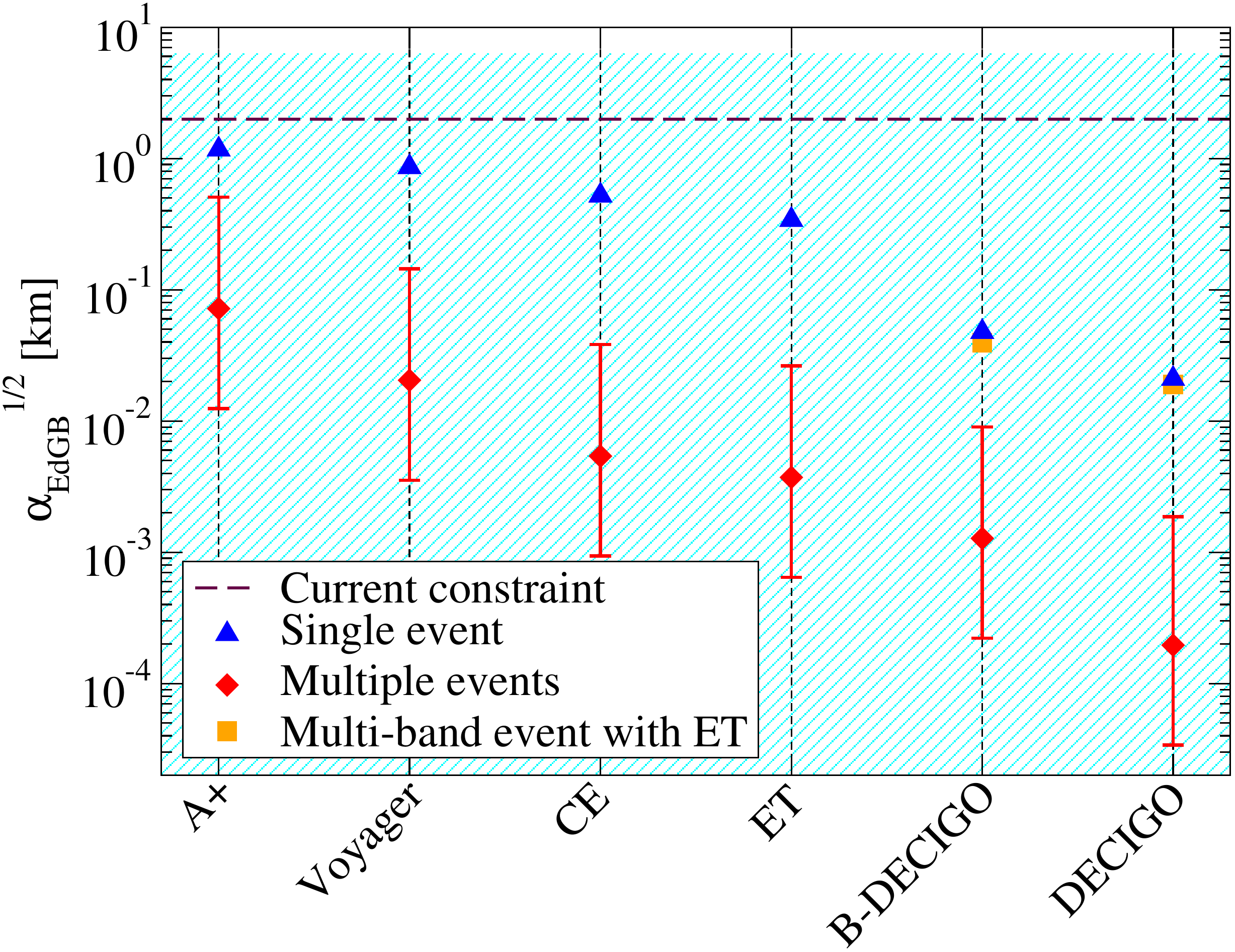}
\caption{
Estimated $68$\% confidence interval constraints on the EdGB coupling parameter $\sqrt{\alpha_\EdGB}$ for a $10\text{ M}_\odot-1.4\text{ M}_\odot$ BH-NS merger event as observed on each detector.
The blue triangles represent single-event detections, while the red error bars correspond to the combined constraints from multiple events, with the upper, central, and lower bounds corresponding to the optimistic, ``realistic", and pessimistic number of detections~\cite{Abadie:2010cf}.
The orange squares give the multi-band result in conjunction with ET, and the shaded cyan region is where the small coupling approximation is valid.
Finally, the horizontal dashed line corresponds to the current most stringent result~\cite{Yagi_EdGB,Nair_dCSMap,Yamada:2019zrb}.
}\label{fig:stackedEDGB}
\end{center}
\end{figure}

We follow this up with a discussion of future constraints placed on EdGB gravity.
Similar to the previous section, we estimate the constraints placed on $\sqrt{\alpha_\EdGB}$ from a $10\text{ M}_\odot-1.4\text{ M}_\odot$ BH-NS merger event detected on each detector, in the single-event, multiple-event, and multi-band cases.
Figure~\ref{fig:stackedEDGB} displays the corresponding bounds for each scenario.
Observe that the single-event rates can place constraints between $0.02-1$ km, all stronger than the current bound of $2$ km.
Further, we see that the multi-band constraints do not offer much improvement from the single-band case, while the combined event bounds can reach down to $\sim 10^{-5}$ km with DECIGO, improving the current bounds by up to five orders-of-magnitude. 
These bounds with DECIGO are consistent with the rough estimate presented in~\cite{Yagi_EdGB} and a recent analysis of~\cite{Gnocchi:2019jzp} for binary black holes with single events.

\section{Validity of Approximations}\label{sec:approximations}
In this section we explore the simplifying approximations made in the above analysis, and their effects on the presented results.
We begin with a discussion on the number of higher-order PN corrections added to the gravitational waveform.
In the main analysis, only the leading order $-1$PN correction term was taken into account in the gravitational waveform, and here we discuss the effect of including corrections up to $0.5$PN order. 
Next we consider the effect of rotating BHs, rather than the static ones considered in the analysis.
Finally we conclude with an alternative method to combine multiple events with varying BH and NS masses, rather than the fixed masses considered in the main analysis.

\subsection{Higher-order STT corrections to the waveform}\label{sec:highorder}
Let us begin with a discussion of the higher-order PN corrections present in the gravitational waveform.
We start by taking into account the $l=m=2$ dipolar contributions to the Fourier domain phase found in Eq.~(81c) of Ref.~\cite{Sennett:2016klh}\footnote{Note the non-dipolar phase corrections are not relative to the $-1$PN phase correction, proportional to $\Delta\alpha$, so were not included in this approximation for simplicity.} up to $0.5$PN order. 
We make the approximation $\alpha_\text{A} \ll 1$ such that only terms proportional to $\Delta\alpha^2$ (and thus also proportional to $\beta_\DEF$) remain, making it possible to keep correlations minimal by allowing only one non-GR parameter to remain in the waveform, $\beta_\DEF$\footnote{Note that the addition of new parameters to the gravitational waveform template may act to weaken the obtained constraints on $\Delta\alpha$ due to increased correlations between the parameters. A detailed analysis on the magnitude of this effect is beyond the scope of this analysis. The error due to linearization of the dipole radiation is expected to be minimal because $\beta_\DEF \ll (\Delta\alpha)^2$.}.
The resulting DEF corrections to the gravitational waveform up to $0.5$PN order are given by
\begin{equation}
  \delta\psi=\beta_\DEF u^{-7} \left(1+\frac{2623+2640\eta}{4280} u^{2}-6\pi u^{3}\right)\,,
\end{equation}
where $\beta_\DEF$ is the $-1$PN ppE parameter given in Eq.~\eqref{eq:ppeBeta}.
We now include these additional corrections into the gravitational waveform and recompute bounds on the DEF theory for an SNR 10 event with a BH mass of $10\text{ M}_\odot$ on the LIGO O3 detector.
We find constraints on $\Delta\alpha$ to be $\sim 0.104$ with leading-order $-1$PN corrections included, and $\sim 0.105$ (negligible difference on a plot such as Fig.~\ref{fig:aligoDEF}) with higher-order corrections to $0.5$PN included.
Such results have a $\sim 0.4$\% difference, and are likely to be negligible in the presented analysis. 
A similar result was found in~\cite{Yunes:2016jcc} for Brans-Dicke theory.
We also bring to attention Ref.~\cite{Tahura:2019dgr}, where it was discovered that the additional presence of non-GR amplitude corrections to the waveform only differs from those in phase by $\sim4$\% - another negligible difference in our analysis.

\subsection{Rotating BHs}\label{sec:rotating}
Now let us discuss the effect of considering rotating BHs in our analysis, rather than the static ones considered previously. 
To perform this simple comparison, we compute constraints on $\Delta\alpha$ in the DEF theory of gravity, and $\sqrt{\alpha_\EdGB}$ in the EdGB theory of gravity for a SNR 10 event with a BH mass of $10\text{ M}_\odot$ on the LIGO O3 detector, with the assumptions of BH spins $\chi_\BH=(0,0.5,1)$.
The resulting constraints were found to be $\Delta\alpha < (0.105,0.104,0.103)$ and $\sqrt{\alpha_\EdGB} < (1.80\text{ km},1.79\text{ km},1.78\text{ km})$ respectively, with only a maximal difference of $\sim1.5$\% found between them in either case.
Thus, we conclude that the effect of rotating BHs in our analysis is sufficiently negligible.

We also point out that the inclusion of spin precession does not play a crucial role in the order-of-magnitude constraint of $-1$PN effects. 
This is demonstrated in Table~IV of Ref.~\cite{Yagi:2009zm}, where it was shown that spin precession strengthens constraints on the Brans-Dicke parameter by $\sim43$\%, due to small correlations between the $-1$PN ppE parameter and the spins entering at $1.5$PN.
Such an effect is magnified for increasing mass-ratio systems, much larger than those considered in this analysis, as they tend to increase the effects of precession.

\subsection{BH/NS mass populations}\label{sec:varymass}
In this section we model an appropriate BH-NS mass distribution function and implement it into the procedure used to combine the uncertainty in non-GR parameters from $N$ events, now with variational masses.
To do this, we modify the expression given in Eq.~\eqref{eq:stacking} by injecting a mass-distribution function $f(m_\BH,m_\NS)$ like so
\begin{align}
    \sigma_{\theta^a}^{-2}=\Delta T\int\limits^{40\text{ M}_\odot}_{5\text{ M}_\odot}\int\limits^{2\text{ M}_\odot}_{1\text{ M}_\odot}\int\limits_0^{z_h(m_\NS, m_\BH)}4\pi \lbrack a_0 r(z) \rbrack^2 \mathcal{R} R(z) &\frac{d\tau}{dz} \sigma_{\theta^a}(z,m_\BH,m_\NS)^{-2}\\
    &\times f(m_\BH,m_\NS) dz dm_\NS dm_\BH,\nonumber
\end{align}
where $\sigma_{\theta^a}(z,m_\BH,m_\NS)$ and $z_h(m_\NS,m_\BH)$ are now interpolated functions that also depend on the individual binary masses. For simplicity, we assume that $f(m_\BH,m_\NS)$ for BH/NS is simply given by a product of the individual mass distributions $f_\BH(m_\BH)$ and $f_\NS(m_\NS)$ as $f(m_\BH,m_\NS)=\mathcal{C}f_\BH(m_\BH)f_\NS(m_\NS)$, where the constant $\mathcal{C}$ is determined by 
normalizing the function to be unity when being integrated over $m_\BH$ and $m_\NS$. We use $f_\BH(m_\BH)$ as the mass distribution of primary black holes in stellar-mass BH binaries derived by the LIGO/Virgo Collaborations~\cite{LIGOScientific:2018jsj}, while we adopt $f_\NS(m_\NS)$ as a Gaussian distribution used e.g. in~\cite{Taylor:2011fs}. Namely, we have
\begin{equation}
    f_\BH(m_\BH) \propto\left\{\begin{array}{ll}{\left(\frac{m_\BH}{\text{M}_\odot}\right)^{-\alpha} } & {\text { if } m_{\min }  \leq m_\BH \leq m_{\max }} \\ {0} & {\text { otherwise }}\end{array}\right\},\hspace{5mm}f_\NS(m_\NS) \propto\mathcal{N}(\mu_\NS,\sigma_\NS).
\end{equation}
The relevant parameters $\alpha$, $m_\text{min}$, $m_\text{max}$, $\mu_\NS$, and $\sigma_\NS$ have been fit to be $0.4$, $5\text{ M}_\odot$, $41.6\text{ M}_\odot$, $1.34\text{ M}_\odot$, and $0.06\text{ M}_\odot$ respectively.
We perform a grid search with 8 redshift values between 0 and 8, 10 NS masses between $1\text{ M}_\odot$ and $2\text{ M}_\odot$ and 10 BH masses between $5\text{ M}_\odot$ and $40\text{ M}_\odot$ to compute $\sigma_{\sqrt{\alpha_\EdGB}}(z,m_\BH,m_\NS)$ for 800 mass/redshift samples, which is then interpolated.

We perform this example computation for the EdGB theory of gravity on the CE detector with the ``realistic" number of events ($730,000$) and compare the resulting constraint on $\sqrt{\alpha_\EdGB}$ to the case of fixed-mass binaries presented in the main analysis.
Under the described circumstances, we find a constraint on $\sqrt{\alpha_\EdGB}$ of $0.003$ km with the new variational mass model. 
Compared to the static-mass model result of $0.004$ km, we find that the two methods agree to within 25\%.
Interestingly, we find the new constraints to be stronger than the old ones as the new analysis includes BH masses lower than $10M_
\odot$ considered originally, and thus the results displayed in the main text can be presented as a conservative estimate.
Because the relationship between $\alpha_0$ and $\beta_0$ in DEF theory of gravity themselves depend on the constituent masses, this method can not be used to reliably compute bounds in the $(\alpha_0,\beta_0)$ plane.
However, we expect to find similar results to the EdGB case (where $\alpha_\EdGB^2$ doesn't depend on the masses). This is verified by instead estimating constraints on $\Delta\alpha$ only, which agrees to the static-mass model to within 25\% as well.


\section{Conclusion}\label{sec:conclusion}
In this analysis, we demonstrated the present and future considerations on constraining STTs which violate the SEP.
We considered both the DEF and EdGB theories, which predict massless scalar fields $\varphi$ which couple to matter and alter the consequent trajectories of gravitating bodies.
We investigate constraints placed on these theories' coupling parameter spaces for the possible detection of BH-NS coalescences, both on the current iteration of LIGO interferometers, and with future GW detectors both on the ground and in space.
In the DEF theory, we find that if such an event (such as the possible candidates S190426c or S190814bv in the O3 run) were to be observed with the present GW detection capabilities, competing bounds to those from pulsar timing observations can be presented.
In EdGB theory, we find that with BH masses less than $19.5\text{ M}_\odot$, improvements to the current constraint on the coupling parameter $\sqrt{\alpha_\EdGB}<2$ km can be made to the order of $\mathcal{O}(0.1)$ km.
Such events detected on future GW detectors (single-event, multi-band observations, and multiple-event stacking) have been demonstrated to improve upon the current bounds by several orders of magnitude in many cases.

Future work in this direction can improve upon this analysis by considering a Bayesian approach to parameter estimation, rather than the Fisher one considered here.
Further, more accurate BH-NS population simulations other than those found in Ref.~\cite{Abadie:2010cf} may be utilized in future analyses, together with different masses for different events.
Finally, one could consider a more comprehensive list of STTs to study, rather than the select few examples investigated here: DEF, MO (see~\ref{app:theoryCompare} for a comparison between the two), and EdGB. One could even consider theories other than STTs, such as those involving vector fields and/or additional tensor fields.


\section{Acknowledgments}\label{sec:acknowledgements} 
We thank Takahiro Tanaka for carefully reading over the work.
We also thank Benjamin Lackey for generously providing us tabulated data for the WFF1, APR4, and MPA1 EoSs used in ~\cite{Read2009}.
Z.C. and K.Y. acknowledges support from NSF Award PHY-1806776. 
K.Y. would like to also acknowledge support by the COST Action GWverse CA16104 and JSPS KAKENHI Grants No. JP17H06358.
B.C.S. and K.Y. acknowledge support from the Mead Endowment.


\appendix
\section{Quasi-Brans-Dicke theory comparison}\label{app:theoryCompare}
\begin{figure}
\begin{center}
\includegraphics[width=.7\columnwidth]{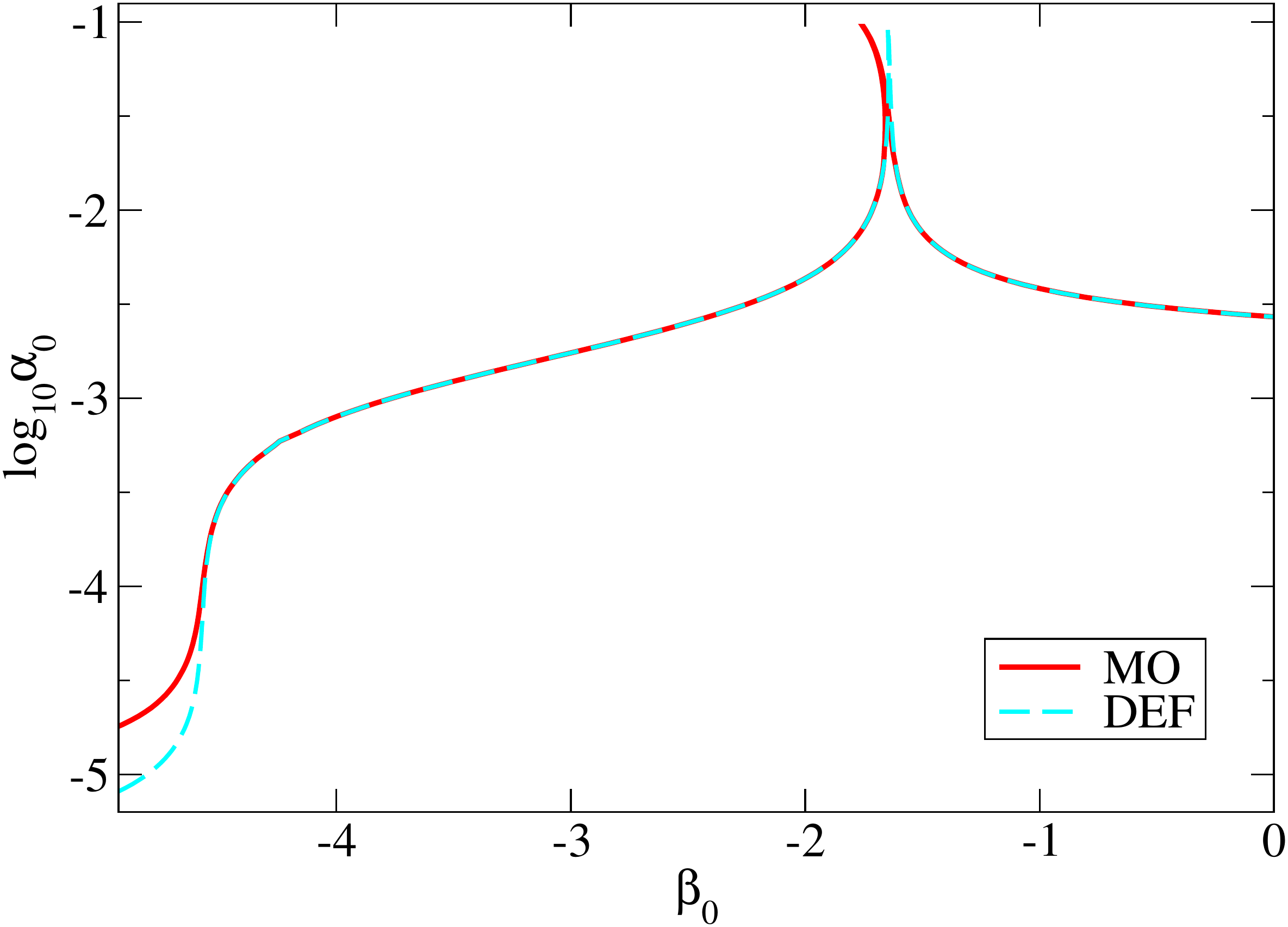}
\caption{
Comparison between the PSR J0337~\cite{Ransom:2014xla,Archibald:2018oxs} constraints formed in the $\alpha_0-\beta_0$ plane for the two quasi-Brans-Dicke theories: DEF and MO.
Bounds formed from GW constraints of BH-NS binaries were found to be indistinguishable from one another in each theory, and have been excluded from this figure.
Observe how for a majority of the contours each theory predicts identical constraints.
The two obvious exceptions being the tilt of the horn, and the drop-off at the lower-left region of the parameter space. 
}\label{fig:theoryCompare}
\end{center}
\end{figure}

In this appendix, we compare results between the DEF~\cite{Damour_1992,Damour_1993} and MO~\cite{Mendes:2016fby} quasi-Brans-Dicke theories of gravity.
The latter theory is defined by the coupling $\alpha(\varphi)=\text{tanh}(\sqrt{3}\beta_0\varphi)/\sqrt{3}$, while the former relies upon only the first term in the above expansion about $\varphi_0$, namely $\alpha(\varphi)=\beta_0 \varphi$.
Figure~\ref{fig:theoryCompare} compares the results for the PSR J0337~\cite{Ransom:2014xla,Archibald:2018oxs} system from the SEP-violation test, assuming an APR4 EoS\footnote{The BH-NS system constraints were found to be indistinguishable between theories, due to their lack of the horn structure present only in pulsar-WD binaries.}.
Observe how the ``horn" structure\footnote{The horns arise in pulsar-WD systems because certain values of $\alpha_0$ and $\beta_0$ suppresses the dipole term  and deteriorating the constraints~\cite{Anderson:2019eay}.} is more pronounced in MO theory, and the drop-off in the lower left region is shifted.
Otherwise, the two theories predict nearly-identical values. This finding is consistent with that in~\cite{Anderson:2019hio} for the orbital decay rate measurement of pulsar-WD binaries.

\section{Equation of state comparison and the effects of spin in the waveform}\label{app:EosSpinCompare}
\begin{figure}
\begin{center}
\includegraphics[width=.7\columnwidth]{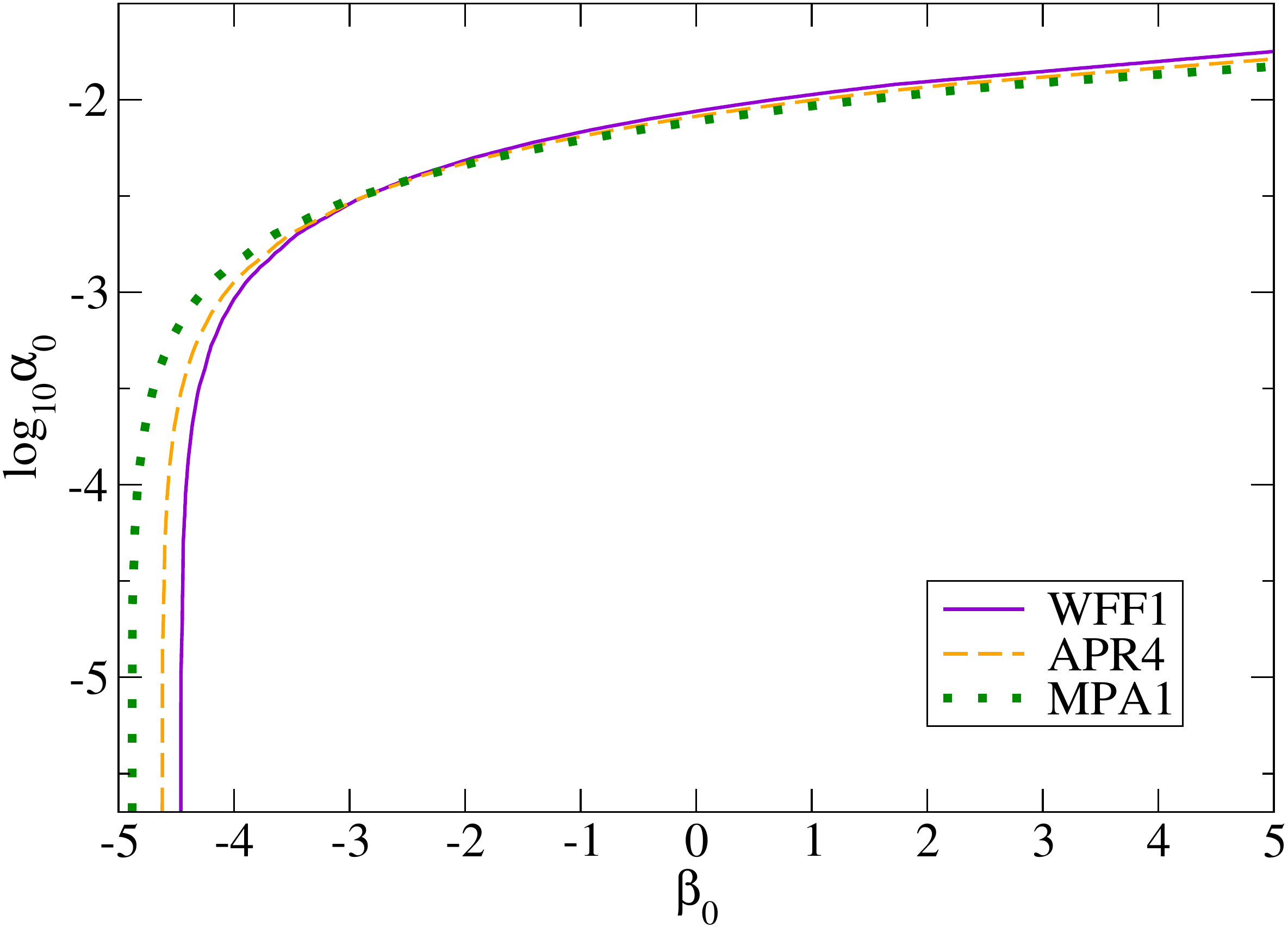}
\caption{Comparison between quasi-Brans-Dicke constraints formed in the $\alpha_0-\beta_0$ plane assuming WFF1, APR4, and MPA1 NS EoSs, all compatible with the observation of GW170817~\cite{LIGO:posterior}.
These constraints were computed assuming a $10\text{ M}_\odot-1.4\text{ M}_\odot$ BH-NS system at 1 Gpc detected by ET.
We observe that softer EoSs give stronger bounds for smaller values of $\beta_0\lessapprox-3$, while stiffer EoSs give stronger constraints for larger values of $\beta_0\gtrapprox-3$.
}\label{fig:eosCompare}
\end{center}
\end{figure}

In this appendix, we present a comparison between the assumption of different NS EoSs, as well as an investigation into spin effects in the gravitational waveform.
Figure~\ref{fig:eosCompare} compares bounds in the $\alpha_0-\beta_0$ plane for $10\text{ M}_\odot-1.4\text{ M}_\odot$ BH-NS system detected by ET, assuming three different EoSs: WFF1~\cite{WFF1,WFF1dat,Read2009}, APR4~\cite{APR4,Read2009}, and MPA1~\cite{MPA1,Read2009}.
Such EoSs were chosen to be consistent with the GW observation of binary NSs, GW170817~\cite{LIGO:posterior,TheLIGOScientific:2017qsa,LIGOScientific:2019eut}, and with increasing degrees of stiffness.
We observe that, while the constraints do not differ much, the softer EoSs produce stronger bounds for small values of $\beta_0\lessapprox-3$, while the stiffer EoSs give stronger results for large values of $\beta_0\gtrapprox-3$.
Thus, for consistency, we present results in the main text for the APR4 EoS.

\begin{figure}
\begin{center}
\includegraphics[width=.7\columnwidth]{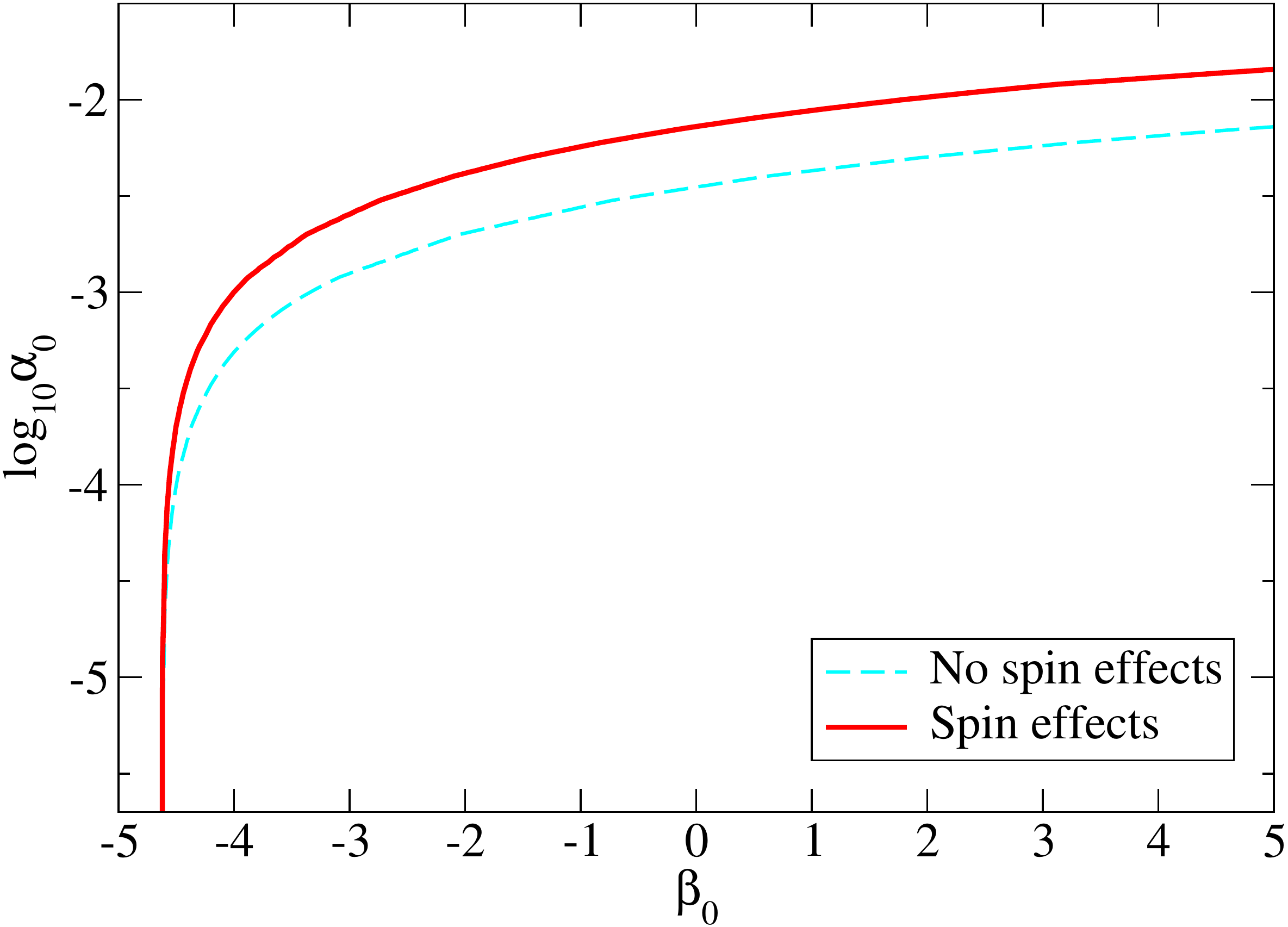}
\caption{Comparison between the quasi-Brans-Dicke constraints formed in the $\alpha_0-\beta_0$ plane with and without including spin effects in the PhenomD gravitational waveform.
The latter, which produces much stronger constraints, can be seen here to under-estimate $\Delta\alpha$ by a factor of $\sim2$.
The constraints displayed here were computed assuming a $10\text{ M}_\odot-1.4\text{ M}_\odot$ BH-NS system at 1 Gpc detected by ET.
}\label{fig:spinCompare}
\end{center}
\end{figure}

We now consider the advisement of including spin effects in the gravitational waveform when computing constraints on quasi-Brans-Dicke theories.
Such bounds were computed for binary NS systems found in Ref.~\cite{Shao:2017gwu} with a waveform template not including any spin effects.
In our analysis, we utilize the PhenomD~\cite{PhenomDII,PhenomDI} gravitational waveform which does indeed include spin effects.
Figure~\ref{fig:spinCompare} compares the constraints formed from the ET observation of a $10\text{ M}_\odot-1.4\text{ M}_\odot$ BH-NS system, both with and without spin effects included in the PhenomD waveform.
We see that the latter under-estimates bounds on $\Delta\alpha$ by a factor of 2, indicating the necessity to include spin effects in the waveform.
These discrepancies arise from the correlations between spin, and the other parameters in the waveform, in particular the non-GR parameter, which ultimately increases the uncertainties in parameter estimation.

\clearpage
\section*{References}
\bibliographystyle{iopart-num}
\bibliography{bibliography.bib}

\newcommand{\noop}[1]{}
\providecommand{\newblock}{}
\begin{thebibliography}{100}
\expandafter\ifx\csname url\endcsname\relax
  \def\url#1{{\tt #1}}\fi
\expandafter\ifx\csname urlprefix\endcsname\relax\def\urlprefix{URL }\fi
\providecommand{\eprint}[2][]{\url{#2}}

\bibitem{Will_SolarSystemTest}
Will C~M 2014 {\em Living Reviews in Relativity\/} {\bf 17} 4 ISSN 1433-8351
  \urlprefix\url{https://doi.org/10.12942/lrr-2014-4}

\bibitem{Stairs_BinaryPulsarTest}
Stairs I~H 2003 {\em Living Rev. Rel.\/} {\bf 6} 5 (\textit{Preprint}
  \eprint{astro-ph/0307536})

\bibitem{Wex_BinaryPulsarTest}
Wex N 2014 {\em to appear in the Brumberg Festschrift, ed. S. M. Kopeikein\/}
  (\textit{Preprint} \eprint{1402.5594})

\bibitem{Ferreira_CosmologyTest}
Ferreira P~G 2019 {\em Annual Review of Astronomy and Astrophysics\/} {\bf 57}
  335--374 (\textit{Preprint} \eprint{"1902.10503"})

\bibitem{Clifton_CosmologyTest}
Clifton T, Ferreira P~G, Padilla A and Skordis C 2012 {\em Phys. Rept.\/} {\bf
  513} 1--189 (\textit{Preprint} \eprint{1106.2476})

\bibitem{Joyce_CosmologyTest}
Joyce A, Jain B, Khoury J and Trodden M 2015 {\em Phys. Rept.\/} {\bf 568}
  1--98 (\textit{Preprint} \eprint{1407.0059})

\bibitem{Koyama_CosmologyTest}
Koyama K 2016 {\em Rept. Prog. Phys.\/} {\bf 79} 046902 (\textit{Preprint}
  \eprint{1504.04623})

\bibitem{Salvatelli_CosmologyTest}
Salvatelli V, Piazza F and Marinoni C 2016 {\em JCAP\/} {\bf 1609} 027
  (\textit{Preprint} \eprint{1602.08283})

\bibitem{Abbott_IMRcon2}
Abbott B~P {\em et~al.\/} (LIGO Scientific, Virgo) 2016 {\em Phys. Rev.
  Lett.\/} {\bf 116} 221101 [Erratum: Phys. Rev. Lett.121,no.12,129902(2018)]
  (\textit{Preprint} \eprint{1602.03841})

\bibitem{Yunes_ModifiedPhysics}
Yunes N, Yagi K and Pretorius F 2016 {\em Phys. Rev. D\/} {\bf 94}(8) 084002
  \urlprefix\url{https://link.aps.org/doi/10.1103/PhysRevD.94.084002}

\bibitem{GW150914}
Abbott B~P {\em et~al.\/} (LIGO Scientific, Virgo) 2016 {\em Phys. Rev.
  Lett.\/} {\bf 116} 241102 (\textit{Preprint} \eprint{1602.03840})

\bibitem{LIGOScientific:2018mvr}
Abbott B~P {\em et~al.\/} (LIGO Scientific, Virgo) 2019 {\em Phys. Rev.\/} {\bf
  X9} 031040 (\textit{Preprint} \eprint{1811.12907})

\bibitem{TheLIGOScientific:2017qsa}
Abbott B~P {\em et~al.\/} (Virgo, LIGO Scientific) 2017 {\em Phys. Rev.
  Lett.\/} {\bf 119} 161101 (\textit{Preprint} \eprint{1710.05832})

\bibitem{Abbott_IMRcon}
Abbott B~P {\em et~al.\/} (The LIGO Scientific Collaboration and the Virgo
  Collaboration) 2019 {\em Phys. Rev. D\/} {\bf 100}(10) 104036
  \urlprefix\url{https://link.aps.org/doi/10.1103/PhysRevD.100.104036}

\bibitem{Monitor:2017mdv}
Abbott B~P {\em et~al.\/} (LIGO Scientific, Virgo, Fermi-GBM, INTEGRAL) 2017
  {\em Astrophys. J.\/} {\bf 848} L13 (\textit{Preprint} \eprint{1710.05834})

\bibitem{Abbott:2018lct}
Abbott B~P {\em et~al.\/} (LIGO Scientific, Virgo) 2019 {\em Phys. Rev.
  Lett.\/} {\bf 123} 011102 (\textit{Preprint} \eprint{1811.00364})

\bibitem{Anderson:2019eay}
Anderson D, Freire P and Yunes N 2019 {\em Class. Quant. Grav.\/} {\bf 36}
  225009 (\textit{Preprint} \eprint{1901.00938})

\bibitem{Freire:2012mg}
Freire P~C~C, Wex N, Esposito-Far\`{e}se G, Verbiest J~P~W, Bailes M, Jacoby
  B~A, Kramer M, Stairs I~H, Antoniadis J and Janssen G~H 2012 {\em Mon. Not.
  Roy. Astron. Soc.\/} {\bf 423} 3328 (\textit{Preprint} \eprint{1205.1450})

\bibitem{Shao:2017gwu}
Shao L, Sennett N, Buonanno A, Kramer M and Wex N 2017 {\em Phys. Rev.\/} {\bf
  X7} 041025 (\textit{Preprint} \eprint{1704.07561})

\bibitem{Berti_ModifiedReviewLarge}
Berti E {\em et~al.\/} 2015 {\em Class. Quant. Grav.\/} {\bf 32} 243001
  (\textit{Preprint} \eprint{1501.07274})

\bibitem{Archibald:2018oxs}
Archibald A~M, Gusinskaia N~V, Hessels J~W~T, Deller A~T, Kaplan D~L, Lorimer
  D~R, Lynch R~S, Ransom S~M and Stairs I~H 2018 {\em Nature\/} {\bf 559}
  73--76 (\textit{Preprint} \eprint{1807.02059})

\bibitem{Bonilla:2019mbm}
Bonilla A, D'Agostino R, Nunes R~C and de~Araujo J~C~N 2019  (\textit{Preprint}
  \eprint{1910.05631})

\bibitem{DAgostino:2019hvh}
D'Agostino R and Nunes R~C 2019 {\em Phys. Rev.\/} {\bf D100} 044041
  (\textit{Preprint} \eprint{1907.05516})

\bibitem{Berti:spaceFreq}
Berti E, Buonanno A and Will C~M 2005 {\em Phys. Rev.\/} {\bf D71} 084025
  (\textit{Preprint} \eprint{gr-qc/0411129})

\bibitem{Takahiro}
Yagi K and Tanaka T 2010 {\em Prog. Theor. Phys.\/} {\bf 123} 1069--1078
  (\textit{Preprint} \eprint{0908.3283})

\bibitem{Sagunski:2017nzb}
Sagunski L, Zhang J, Johnson M~C, Lehner L, Sakellariadou M, Liebling S~L,
  Palenzuela C and Neilsen D 2018 {\em Phys. Rev.\/} {\bf D97} 064016
  (\textit{Preprint} \eprint{1709.06634})

\bibitem{Huang:2018pbu}
Huang J, Johnson M~C, Sagunski L, Sakellariadou M and Zhang J 2019 {\em Phys.
  Rev.\/} {\bf D99} 063013 (\textit{Preprint} \eprint{1807.02133})

\bibitem{gracedb}
{GraceDB} \url{https://gracedb.ligo.org/superevents/public/O3/}

\bibitem{gracedb2}
{GraceDB} \url{https://gcn.gsfc.nasa.gov/gcn3/25549.gcn3}

\bibitem{classification}
{LVC Classification}
  \url{https://emfollow.docs.ligo.org/userguide/content.html#classification-diagram}

\bibitem{Damour:1992we}
Damour T and Esposito-Far\`{e}se G 1992 {\em Class. Quant. Grav.\/} {\bf 9}
  2093--2176

\bibitem{Damour:1996ke}
Damour T and Esposito-Far\`{e}se G 1996 {\em Phys. Rev.\/} {\bf D54} 1474--1491
  (\textit{Preprint} \eprint{gr-qc/9602056})

\bibitem{Damour:1993hw}
Damour T and Esposito-Far\`{e}se G 1993 {\em Phys. Rev. Lett.\/} {\bf 70}
  2220--2223

\bibitem{Kanti_EdGB}
Kanti P, Mavromatos N~E, Rizos J, Tamvakis K and Winstanley E 1996 {\em Phys.
  Rev.\/} {\bf D54} 5049--5058 (\textit{Preprint} \eprint{hep-th/9511071})

\bibitem{Maeda:2009uy}
Maeda K~i, Ohta N and Sasagawa Y 2009 {\em Phys. Rev.\/} {\bf D80} 104032
  (\textit{Preprint} \eprint{0908.4151})

\bibitem{Campbell:1991kz}
Campbell B~A, Kaloper N and Olive K~A 1992 {\em Phys. Lett.\/} {\bf B285}
  199--205

\bibitem{Yunes:2011we}
Yunes N and Stein L~C 2011 {\em Phys. Rev.\/} {\bf D83} 104002
  (\textit{Preprint} \eprint{1101.2921})

\bibitem{Yagi_EdGBmap}
Yagi K, Stein L~C, Yunes N and Tanaka T 2012 {\em Phys. Rev.\/} {\bf D85}
  064022 [Erratum: Phys. Rev.D93,no.2,029902(2016)] (\textit{Preprint}
  \eprint{1110.5950})

\bibitem{Sotiriou:2014pfa}
Sotiriou T~P and Zhou S~Y 2014 {\em Phys. Rev.\/} {\bf D90} 124063
  (\textit{Preprint} \eprint{1408.1698})

\bibitem{Yagi:2015oca}
Yagi K, Stein L~C and Yunes N 2016 {\em Phys. Rev.\/} {\bf D93} 024010
  (\textit{Preprint} \eprint{1510.02152})

\bibitem{Barausse:2016eii}
Barausse E, Yunes N and Chamberlain K 2016 {\em Phys. Rev. Lett.\/} {\bf 116}
  241104 (\textit{Preprint} \eprint{1603.04075})

\bibitem{Carson_multiBandPRL}
Carson Z and Yagi K 2019  (\textit{Preprint} \eprint{1905.13155})

\bibitem{Nair:2015bga}
Nair R, Jhingan S and Tanaka T 2016 {\em PTEP\/} {\bf 2016} 053E01
  (\textit{Preprint} \eprint{1504.04108})

\bibitem{Nair:2018bxj}
Nair R and Tanaka T 2018 {\em JCAP\/} {\bf 1808} 033 [Erratum:
  JCAP1811,no.11,E01(2018)] (\textit{Preprint} \eprint{1805.08070})

\bibitem{Abadie:2010cf}
Abadie J {\em et~al.\/} (LIGO Scientific, VIRGO) 2010 {\em Class. Quant.
  Grav.\/} {\bf 27} 173001 (\textit{Preprint} \eprint{1003.2480})

\bibitem{Yunes:2009ke}
Yunes N and Pretorius F 2009 {\em Phys. Rev.\/} {\bf D80} 122003
  (\textit{Preprint} \eprint{0909.3328})

\bibitem{Tahura_GdotMap}
Tahura S and Yagi K 2018 {\em Phys. Rev.\/} {\bf D98} 084042 (\textit{Preprint}
  \eprint{1809.00259})

\bibitem{aLIGO}
Advanced {LIGO} \url{https://www.advancedligo.mit.edu/} accessed: 2019-01-10
  \urlprefix\url{https://www.advancedligo.mit,.edu/}

\bibitem{O3}
Ligo gitlab
  \url{https://git.ligo.org/lscsoft/lalsuite/blob/master/lalsimulation/src/LIGO-T1800545-v1-aLIGO_140Mpc.txt}
  accessed: 2019-11-04
  \urlprefix\url{https://git.ligo.org/lscsoft/lalsuite/blob/master/lalsimulation/lib/LIGO-T1800545-v1-aLIGO_140Mpc.txt}

\bibitem{AppData}
Ligo gitlab
  \url{https://git.ligo.org/lscsoft/lalsuite/blob/master/lalsimulation/lib/LIGO-T1800042-v5-aLIGO_APLUS.txt}
  accessed: 2019-11-04
  \urlprefix\url{https://git.ligo.org/lscsoft/lalsuite/blob/master/lalsimulation/src/LIGO-T1800042-v5-aLIGO_APLUS.txt}

\bibitem{ApVoyagerCE}
Ligo-t1400316-v4: Instrument science white paper
  \url{https://dcc.ligo.org/ligo-T1400316/public} accessed: 2019-01-10
  \urlprefix\url{https://dcc.ligo.org/ligo-T1400316/public}

\bibitem{VRTCEETData}
Ligo gitlab https://git.ligo.org/evan.hall/gw-horizon-plot/tree/master/data
  accessed: 2019-11-04
  \urlprefix\url{https://git.ligo.org/evan.hall/gw-horizon-plot/tree/master/data}

\bibitem{Evans:2016mbw}
Abbott B~P {\em et~al.\/} (LIGO Scientific) 2017 {\em Class. Quant. Grav.\/}
  {\bf 34} 044001 (\textit{Preprint} \eprint{1607.08697})

\bibitem{B-DECIGO}
Isoyama S, Nakano H and Nakamura T 2018 {\em PTEP\/} {\bf 2018} 073E01
  (\textit{Preprint} \eprint{1802.06977})

\bibitem{DECIGO}
Yagi K and Seto N 2011 {\em Phys. Rev.\/} {\bf D83} 044011 [Erratum: Phys.
  Rev.D95,no.10,109901(2017)] (\textit{Preprint} \eprint{1101.3940})

\bibitem{PhenomDII}
Husa S, Khan S, Hannam M, P\"urrer M, Ohme F, Forteza X~J and Boh\'e A 2016
  {\em Phys. Rev. D\/} {\bf 93}(4) 044006
  \urlprefix\url{https://link.aps.org/doi/10.1103/PhysRevD.93.044006}

\bibitem{PhenomDI}
Khan S, Husa S, Hannam M, Ohme F, P\"urrer M, Forteza X~J and Boh\'e A 2016
  {\em Phys. Rev. D\/} {\bf 93}(4) 044007
  \urlprefix\url{https://link.aps.org/doi/10.1103/PhysRevD.93.044007}

\bibitem{Lackey:2013axa}
Lackey B~D, Kyutoku K, Shibata M, Brady P~R and Friedman J~L 2014 {\em Phys.
  Rev.\/} {\bf D89} 043009 (\textit{Preprint} \eprint{1303.6298})

\bibitem{Kumar:2016zlj}
Kumar P, P{\"u}rrer M and Pfeiffer H~P 2017 {\em Phys. Rev.\/} {\bf D95} 044039
  (\textit{Preprint} \eprint{1610.06155})

\bibitem{Pannarale:2015jka}
Pannarale F, Berti E, Kyutoku K, Lackey B~D and Shibata M 2015 {\em Phys.
  Rev.\/} {\bf D92} 084050 (\textit{Preprint} \eprint{1509.00512})

\bibitem{Hinderer:2016eia}
Hinderer T {\em et~al.\/} 2016 {\em Phys. Rev. Lett.\/} {\bf 116} 181101
  (\textit{Preprint} \eprint{1602.00599})

\bibitem{Barkett:2019tus}
Barkett K, Chen Y, Scheel M~A and Varma V 2019  (\textit{Preprint}
  \eprint{1911.10440})

\bibitem{Chakravarti:2018uyi}
Chakravarti K {\em et~al.\/} 2019 {\em Phys. Rev.\/} {\bf D99} 024049
  (\textit{Preprint} \eprint{1809.04349})

\bibitem{Wade:tidalCorrections}
Wade L, Creighton J~D~E, Ochsner E, Lackey B~D, Farr B~F, Littenberg T~B and
  Raymond V 2014 {\em Phys. Rev. D\/} {\bf 89}(10) 103012
  \urlprefix\url{https://link.aps.org/doi/10.1103/PhysRevD.89.103012}

\bibitem{Cutler:Fisher}
Cutler C and Flanagan E~E 1994 {\em Phys. Rev. D\/} {\bf 49}(6) 2658--2697
  \urlprefix\url{https://link.aps.org/doi/10.1103/PhysRevD.49.2658}

\bibitem{damour-nagar}
Damour T and Nagar A 2009 {\em Phys.Rev.\/} {\bf D80} 084035 (\textit{Preprint}
  \eprint{0906.0096})

\bibitem{Binnington:2009bb}
Binnington T and Poisson E 2009 {\em Phys. Rev.\/} {\bf D80} 084018
  (\textit{Preprint} \eprint{0906.1366})

\bibitem{Kol:2011vg}
Kol B and Smolkin M 2012 {\em JHEP\/} {\bf 02} 010 (\textit{Preprint}
  \eprint{1110.3764})

\bibitem{Chakrabarti:2013lua}
Chakrabarti S, Delsate T and Steinhoff J 2013  (\textit{Preprint}
  \eprint{1304.2228})

\bibitem{Gurlebeck:2015xpa}
G{\"u}rlebeck N 2015 {\em Phys. Rev. Lett.\/} {\bf 114} 151102
  (\textit{Preprint} \eprint{1503.03240})

\bibitem{Gralla:2017djj}
Gralla S~E 2018 {\em Class. Quant. Grav.\/} {\bf 35} 085002 (\textit{Preprint}
  \eprint{1710.11096})

\bibitem{Harry:2018hke}
Harry I and Hinderer T 2018 {\em Class. Quant. Grav.\/} {\bf 35} 145010
  (\textit{Preprint} \eprint{1801.09972})

\bibitem{LIGO:posterior}
Abbott B~P {\em et~al.\/} (LIGO Scientific, Virgo) 2018 {\em Phys. Rev.
  Lett.\/} {\bf 121} 161101 (\textit{Preprint} \eprint{1805.11581})

\bibitem{Yunes:2016jcc}
Yunes N, Yagi K and Pretorius F 2016 {\em Phys. Rev.\/} {\bf D94} 084002
  (\textit{Preprint} \eprint{1603.08955})

\bibitem{Anderson:2019hio}
Anderson D and Yunes N 2019 {\em Class. Quant. Grav.\/} {\bf 36} 165003
  (\textit{Preprint} \eprint{1901.00937})

\bibitem{Mendes:2016fby}
Mendes R~F~P and Ortiz N 2016 {\em Phys. Rev.\/} {\bf D93} 124035
  (\textit{Preprint} \eprint{1604.04175})

\bibitem{Damour:1992kf}
Damour T and Nordtvedt K 1993 {\em Phys. Rev. Lett.\/} {\bf 70} 2217--2219

\bibitem{Sampson2014}
Sampson L, Yunes N, Cornish N, Ponce M, Barausse E, Klein A, Palenzuela C and
  Lehner L 2014 {\em Phys. Rev. D\/} {\bf 90}(12) 124091
  \urlprefix\url{https://link.aps.org/doi/10.1103/PhysRevD.90.124091}

\bibitem{Anderson:2016aoi}
Anderson D, Yunes N and Barausse E 2016 {\em Phys. Rev.\/} {\bf D94} 104064
  (\textit{Preprint} \eprint{1607.08888})

\bibitem{Anson:2019ebp}
Anson T, Babichev E and Ramazanov S 2019  (\textit{Preprint}
  \eprint{1905.10393})

\bibitem{Damour:1998jk}
Damour T and Esposito-Far\`{e}se G 1998 {\em Phys. Rev.\/} {\bf D58} 042001
  (\textit{Preprint} \eprint{gr-qc/9803031})

\bibitem{Littenberg:2018xxx}
Littenberg T~B and Yunes N 2019 {\em Class. Quant. Grav.\/} {\bf 36} 095017
  (\textit{Preprint} \eprint{1811.01093})

\bibitem{Zhao:2019suc}
Zhao J, Shao L, Cao Z and Ma B~Q 2019 {\em Phys. Rev.\/} {\bf D100} 064034
  (\textit{Preprint} \eprint{1907.00780})

\bibitem{Abbott:LTposterior}
Abbott B~P {\em et~al.\/} (LIGO Scientific, Virgo) 2019 {\em Phys. Rev.\/} {\bf
  X9} 011001 (\textit{Preprint} \eprint{1805.11579})

\bibitem{Ransom:2014xla}
Ransom S~M {\em et~al.\/} 2014 {\em Nature\/} {\bf 505} 520 (\textit{Preprint}
  \eprint{1401.0535})

\bibitem{Kilic:2014yxa}
Kilic M, Hermes J~J, Gianninas A and Brown W~R 2015 {\em Mon. Not. Roy. Astron.
  Soc.\/} {\bf 446} L26--L30 (\textit{Preprint} \eprint{1410.4898})

\bibitem{Antoniadis:2013pzd}
Antoniadis J {\em et~al.\/} 2013 {\em Science\/} {\bf 340} 6131
  (\textit{Preprint} \eprint{1304.6875})

\bibitem{Bertotti:2003rm}
Bertotti B, Iess L and Tortora P 2003 {\em Nature\/} {\bf 425} 374--376

\bibitem{mignard_2009}
Mignard F and Klioner S~A 2009 {\em Proceedings of the International
  Astronomical Union\/} {\bf 5}(S261) 306
  \urlprefix\url{https://doi.org/10.1017/S174392130999055X}

\bibitem{Carson_multiBandPRD}
Carson Z and Yagi K \noop{3001}in preparation

\bibitem{Gnocchi:2019jzp}
Gnocchi G, Maselli A, Abdelsalhin T, Giacobbo N and Mapelli M 2019 {\em Phys.
  Rev.\/} {\bf D100} 064024 (\textit{Preprint} \eprint{1905.13460})

\bibitem{Moore:2019pke}
Moore C~J, Gerosa D and Klein A 2019 {\em Mon. Not. Roy. Astron. Soc.\/} {\bf
  488} L94--L98 (\textit{Preprint} \eprint{1905.11998})

\bibitem{Zhang:2017sym}
Zhang X, Yu J, Liu T, Zhao W and Wang A 2017 {\em Phys. Rev.\/} {\bf D95}
  124008 (\textit{Preprint} \eprint{1703.09853})

\bibitem{Sotiriou:2013qea}
Sotiriou T~P and Zhou S~Y 2014 {\em Phys. Rev. Lett.\/} {\bf 112} 251102
  (\textit{Preprint} \eprint{1312.3622})

\bibitem{Bakopoulos:2018nui}
Bakopoulos A, Antoniou G and Kanti P 2019 {\em Phys. Rev.\/} {\bf D99} 064003
  (\textit{Preprint} \eprint{1812.06941})

\bibitem{Antoniou:2017hxj}
Antoniou G, Bakopoulos A and Kanti P 2018 {\em Phys. Rev.\/} {\bf D97} 084037
  (\textit{Preprint} \eprint{1711.07431})

\bibitem{Antoniou:2017acq}
Antoniou G, Bakopoulos A and Kanti P 2018 {\em Phys. Rev. Lett.\/} {\bf 120}
  131102 (\textit{Preprint} \eprint{1711.03390})

\bibitem{Yagi_EdGB}
Yagi K 2012 {\em Phys. Rev.\/} {\bf D86} 081504 (\textit{Preprint}
  \eprint{1204.4524})

\bibitem{Nair_dCSMap}
Nair R, Perkins S, Silva H~O and Yunes N 2019 {\em Phys. Rev. Lett.\/} {\bf
  123} 191101 (\textit{Preprint} \eprint{1905.00870})

\bibitem{Yamada:2019zrb}
Yamada K, Narikawa T and Tanaka T 2019 {\em PTEP\/} {\bf 2019} 103E01
  (\textit{Preprint} \eprint{1905.11859})

\bibitem{Lattimer:2019qdc}
Lattimer J~M 2019  (\textit{Preprint} \eprint{1908.03622})

\bibitem{Sennett:2016klh}
Sennett N, Marsat S and Buonanno A 2016 {\em Phys. Rev.\/} {\bf D94} 084003
  (\textit{Preprint} \eprint{1607.01420})

\bibitem{Tahura:2019dgr}
Tahura S, Yagi K and Carson Z 2019 {\em Phys. Rev.\/} {\bf D100} 104001
  (\textit{Preprint} \eprint{1907.10059})

\bibitem{Yagi:2009zm}
Yagi K and Tanaka T 2010 {\em Phys. Rev.\/} {\bf D81} 064008 [Erratum: Phys.
  Rev.D81,109902(2010)] (\textit{Preprint} \eprint{0906.4269})

\bibitem{LIGOScientific:2018jsj}
Abbott B~P {\em et~al.\/} (LIGO Scientific, Virgo) 2019 {\em Astrophys. J.\/}
  {\bf 882} L24 (\textit{Preprint} \eprint{1811.12940})

\bibitem{Taylor:2011fs}
Taylor S~R, Gair J~R and Mandel I 2012 {\em Phys. Rev.\/} {\bf D85} 023535
  (\textit{Preprint} \eprint{1108.5161})

\bibitem{Read2009}
Read J~S, Lackey B~D, Owen B~J and Friedman J~L 2009 {\em Physical Review D\/}
  {\bf 79} \urlprefix\url{https://doi.org/10.1103/physrevd.79.124032}

\bibitem{Damour_1992}
Damour T and Esposito-Far\`{e}se G 1992 {\em Classical and Quantum Gravity\/}
  {\bf 9} 2093--2176
  \urlprefix\url{https://doi.org/10.1088%2F0264-9381%2F9%2F9%2F015}

\bibitem{Damour_1993}
Damour T and Esposito-Far\`ese G 1993 {\em Phys. Rev. Lett.\/} {\bf 70}(15)
  2220--2223
  \urlprefix\url{https://link.aps.org/doi/10.1103/PhysRevLett.70.2220}

\bibitem{WFF1}
Wiringa R~B, Fiks V and Fabrocini A 1988 {\em Phys. Rev. C\/} {\bf 38}(2)
  1010--1037 \urlprefix\url{https://link.aps.org/doi/10.1103/PhysRevC.38.1010}

\bibitem{WFF1dat}
Center for gravitation, cosmology \& astrophysics sources
  http://www.gravity.phys.uwm.edu/rns/source/eos/ accessed: 2019-11-05
  \urlprefix\url{http://www.gravity.phys.uwm.edu/rns/source/eos/}

\bibitem{APR4}
Akmal A, Pandharipande V~R and Ravenhall D~G 1998 {\em Phys. Rev. C\/} {\bf
  58}(3) 1804--1828
  \urlprefix\url{https://link.aps.org/doi/10.1103/PhysRevC.58.1804}

\bibitem{MPA1}
M{\"u}ther H, Prakash M and Ainsworth T~L 1987 {\em Phys. Lett.\/} {\bf B199}
  469--474

\bibitem{LIGOScientific:2019eut}
Abbott B~P {\em et~al.\/} (LIGO Scientific, Virgo) 2019  (\textit{Preprint}
  \eprint{1908.01012})

\end{thebibliography}

\end{document}